\documentclass[prb,twocolumn,superscriptaddress]{revtex4-1}
\usepackage[latin1]{inputenc}
\usepackage{graphicx}
\usepackage{amssymb}
\usepackage{amsmath}
\usepackage{xspace}
\usepackage{dcolumn}
\usepackage{bm}
\usepackage{color}
\usepackage{float}
\usepackage[extra]{tipa}
\usepackage{mathtools}

\setcitestyle{square,numbers,compress}

\newfont{\tensy}{cmsy10}

\newcommand{\ie}[0]{i.e.\@\xspace}
\newcommand{\eg}[0]{e.g.\@\xspace}
\newcommand{\etal}[0]{\emph{et al.\@\xspace}}

\newcommand{\bk}{\boldsymbol{k}}

\newcommand{\bq}{\boldsymbol{q}}

\newcommand{\RefI}{\cite{PhysRevLett.121.086601}\xspace}

\newcommand{\rmi}{\text{i}}
\newcommand{\UP}[0]{\uparrow}
\newcommand{\DO}[0]{\downarrow}

\newcommand{\oQ}{\hat{Q}}
\newcommand{\oH}{\hat{H}}

\newcommand{\on}{\hat{n}}

\newcommand{\oxi}{\hat{\xi}}

\newcommand{\os}{\hat{s}}
\renewcommand{\oc}{\hat{c}}
\newcommand{\of}{\hat{f}}

\newcommand{\rmd}{\text{d}}

\newcommand{\ZII}{Z$_2$\xspace}

\newcommand{\bit}{\begin{itemize}}
\newcommand{\eit}{\end{itemize}}

\newcommand{\oh}{\mbox{$\frac{1}{2}$}}

\newcommand{\om}[0]{\omega}

\newcommand{\EF}{E_\text{F}}

\newcommand{\kB}{k_\text{B}}
\newcommand{\nag}{{\phantom{\dag}}}

\newcommand{\las}[0]{\langle}
\newcommand{\ras}[0]{\rangle}

\newcommand{\la}[0]{\left\las}
\newcommand{\ra}[0]{\right\ras}
\newcommand{\ket}[1]{\left|#1\ra}
\newcommand{\bra}[1]{\la#1\right|}
\newcommand{\tr}[0]{\text{tr}}

\begin{document}

%%%%%%%%%%%%%%%%%%%%%%%%%%%%%%%%%%%%%%%%%%%%%%%%%%%%%%%%%%%%%%%%%%%%%%%%%%%%%
%%%%%%%%%%%%%%%%%%%%% TITLE & ABSTRACT %%%%%%%%%%%%%%%%%%%%%%%
%%%%%%%%%%%%%%%%%%%%%%%%%%%%%%%%%%%%%%%%%%%%%%%%%%%%%%%%%%%%%%%%%%%%%%%%%%%%%

\title{Orthogonal metal in the Hubbard model with liberated slave spins}

\author{Martin Hohenadler}
\affiliation{\mbox{Institut f\"ur Theoretische Physik und Astrophysik,
    Universit\"at W\"urzburg, 97074 W\"urzburg, Germany}}

\author{Fakher F. Assaad}
\affiliation{\mbox{Institut f\"ur Theoretische Physik und Astrophysik,
    Universit\"at W\"urzburg, 97074 W\"urzburg, Germany}}
\affiliation{W\"urzburg-Dresden Cluster of Excellence ct.qmat, 97074 W\"urzburg, Germany}

\begin{abstract}
A two-dimensional Falicov-Kimball model, equivalent to the Hubbard model in
an unconstrained slave-spin representation, is studied by quantum Monte Carlo simulations.
The focus is on a fractionalized metallic phase that is characterized in terms
of spectral, thermodynamic, and transport properties, including a comparison
to the half-filled Hubbard model. The properties of this phase, most notably 
a single-particle gap but gapless spin and charge excitations, can in
principle be understood in the framework of orthogonal metals.
However, important and interesting differences arise in the present setting
compared to single-particle mean-field theories and other models.
We also discuss the role of the local constraints from the slave-spin
representation within an extended phase diagram that includes the
spatial dimension as a parameter, thereby making contact with previous work
in infinite dimensions. Finally, we highlight the absence of $\pi$-flux
configurations in the slave-spin formulation, in particular in the context of
topologically ordered fractional phases predicted at the mean-field level.
\end{abstract}

\date{\today}

\maketitle

\section{Introduction}\label{sec:introduction}

Strong electronic correlations underlie some of the most fascinating
discoveries of condensed matter physics, including high-temperature
superconductivity \cite{Bednorz86}, the fractional quantum Hall effect
\cite{PhysRevLett.48.1559}, and quantum spin liquids \cite{RevModPhys.89.025003}.
Even in the absence of phase transitions, correlations can
significantly modify physical properties compared to the widely applicable 
Fermi liquid (FL) paradigm. For example, in one-dimensional
(1D) Luttinger liquids, the original
electronic excitations have zero overlap with the collective density and spin
excitations that determine the low-energy properties
\cite{Giamarchi}. Experimental evidence for non-FL physics has been collected
for quasi-2D materials such as the cuprates, including a pseudogap in the
single-particle spectrum and a linear and/or non-saturating
temperature-dependent resistivity
that exceeds the Mott-Ioffe-Regel bound for a quasiparticle description
\cite{hussey2004universality,greene2019strange}. A unified theory of the
linear resistivity remains an important open problem
\cite{zaanen2018planckian}. Finally, emergent degrees of freedom, in particular gauge fields,
play a key role for our understanding \cite{FradkinBook}.

The Hubbard model is perhaps the most widely studied model of correlated
electrons \cite{PhysRevX.5.041041}. At particle-hole symmetric
points, in particular the half-filled square lattice with nearest-neighbor hopping,
powerful quantum Monte Carlo (QMC) methods can be applied. The ground
state is an antiferromagnetic Mott insulator for any nonzero repulsive
interaction $U$ \cite{Hirsch85}. Away from half-filling, recent work has
revealed a linear high-temperature resistivity in finite and infinite dimensions
\cite{PhysRevLett.110.086401,PhysRevLett.114.246402,PhysRevB.94.235115,huang2018strange} as well as stripe order
\cite{zheng2017stripe,huang2017numerical}. A fermionic
quantum critical point separating a semimetal from an antiferromagnet,
described by a  Gross-Neveu-Yukawa theory  \cite{PhysRevLett.97.146401,Herbut09a}, is observed at $U_c>0$ on the
honeycomb lattice \cite{Sorella92,Assaad13,Sorella12,Toldin14,Otsuka16}.  There have also been significant
advances in studying the Hubbard model using cold-atom quantum simulators
\cite{esslinger2010fermi,mazurenko2016experimental,brown2019bad}.

Slave-spin representations of fermions have proven useful to explore
fractionalization at the mean-field level \cite{PhysRevB.72.205124,PhysRevB.81.035106,PhysRevB.81.155118}.
For example, they provide an order parameter for the paramagnetic
Mott-Hubbard transition in infinite dimensions \cite{PhysRevB.91.245130}. Of particular interest
for our work is the observation of Ref.~\cite{PhysRevB.86.045128} that---beyond
single-site mean-field theories---the gapped phase of the slave spins is
not a Mott insulator but a fractionalized {\em orthogonal metal} (OM,
see Sec.~\ref{sec:model:mean-field} for more details).
OMs are characterized by a gap for single-particle excitations but gapless 
transport and may be
considered the simplest example of non-FLs \cite{PhysRevB.86.045128}. They
provide a framework to reconcile the absence of quasiparticle excitations in
ARPES with the presence of a Fermi surface in quantum oscillation measurements
\cite{PhysRevLett.115.146401}. Using the \ZII slave-spin representation, the
concept of orthogonal phases has been extended to, \eg, Dirac semimetals
\cite{zhong2012correlated}. 

\begin{figure}[b]
  \includegraphics[width=0.425\textwidth]{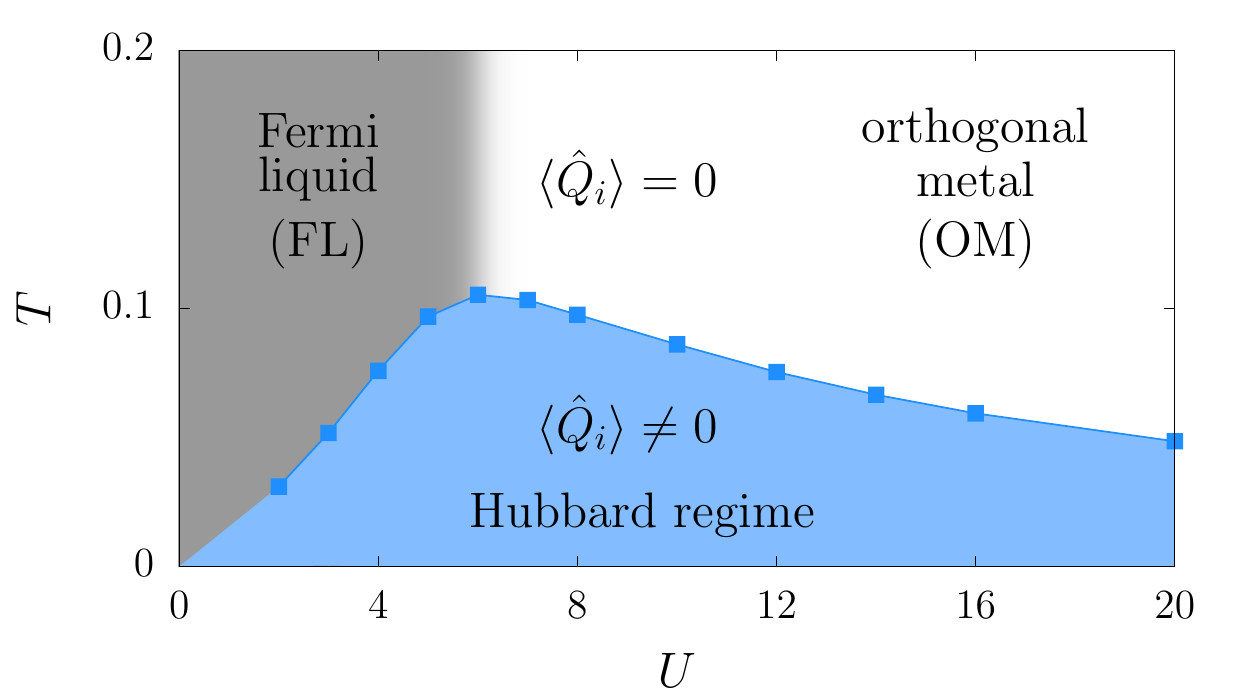}  
  \caption{\label{fig:phasediagram} 
    Phase diagram of the dual Hamiltonians~(\ref{eq:HcQ}),
    (\ref{eq:Hfs}), and~(\ref{eq:Hftau}). A phase
    transition of the Ising variables $\oQ_i$ at $T_Q$ separates the
    low-temperature Hubbard regime from the high-temperature phase. Here, $T_Q$
    was determined from data for $L=8$, see Fig.~\ref{fig:constraints}. The
    two metallic regimes at $T>T_Q$ appear to be separated by a
    crossover, as indicated by the color gradient. Adapted from Ref.~\RefI.}
\end{figure}

\begin{figure}[t]
  \includegraphics[width=0.425\textwidth]{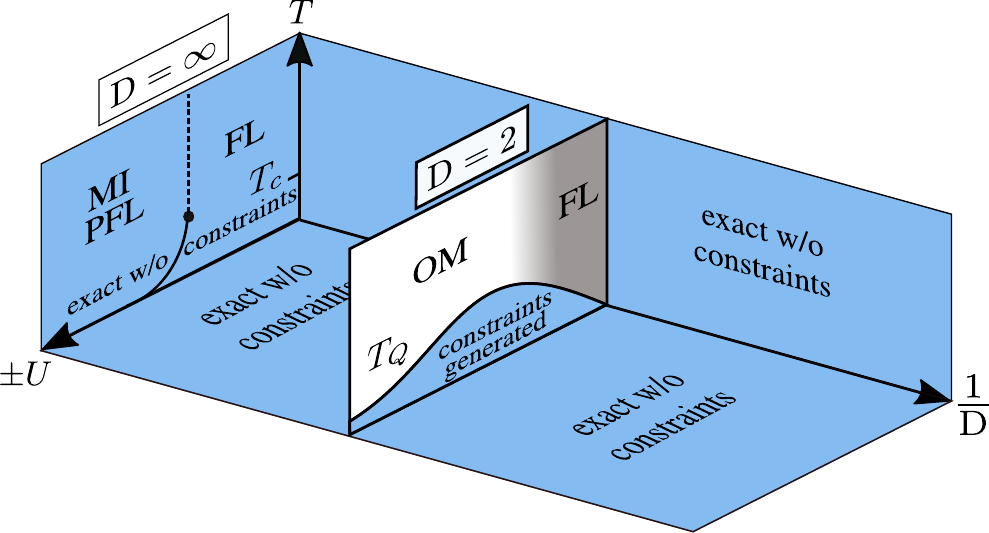}  
  \caption{\label{fig:grandphasediagram}    
    Schematic phase diagram of the Hubbard model in the unconstrained
    slave-spin representation defined by Hamiltonian~(\ref{eq:Hfs})
    [equivalent to Hamiltonian~(\ref{eq:HcQ})]. The axes correspond to
    the interaction $U$, the temperature $T$, and the inverse spatial dimension $1/\text{D}$. In
    the blue-shaded regions, the constraints $\oQ_i$ from the slave-spin
    representation (identical to the local Ising variables in the FKM) are
    either {\em irrelevant or dynamically generated} so that the unconstrained models
    give the same results as the original Hubbard model. For $\text{D}=\infty$,
    paramagnetic DMFT yields a first-order phase transition from an FL to a
    Mott insulator (MI, repulsive case) or a paired Fermi liquid (PFL,
    attractive case) up to $T=T_c$, and a crossover (dashed
     line) for $T>T_c$ \cite{PhysRevLett.101.186403}.}
\end{figure}

In Ref.~\RefI, we introduced a Falicov-Kimball model (FKM) of spinful fermions
with a three-body interaction. It reduces to the Hubbard model at low
temperatures and in infinite dimensions.  QMC
simulations of this model revealed a rather interesting phase diagram, see
Fig.~\ref{fig:phasediagram}. It features a line of thermal Ising phase
transitions with critical temperature $T_Q(U)$, at which the
localized degrees of freedom order ferromagnetically, and a crossover between
two distinct metallic regimes at $T>T_Q$. The weak-coupling regime
($U\lesssim 6$) exhibits qualitatively FL-like properties, with thermodynamic
and transport quantities determined by well-defined quasiparticles. 
This picture breaks down at stronger couplings where metallic
behavior and gapless spin excitations coexist with a single-particle
gap, highly reminiscent of OMs. 

The connection to OMs and fractionalization can be made quite explicit via an
exact duality between the FKM representation and a completely equivalent (\ie,
dual) \ZII slave-spin representation. In fact, our work began as an
investigation of the Hubbard model in the slave-spin representation. 
QMC simulations reveal a clear correlation between the disappearance of
quasiparticles and the disordering of the slave spins. As demonstrated
below, in the low-temperature phase in Fig.~\ref{fig:phasediagram}, labeled
{\em Hubbard regime}, the FKM has the same symmetry as the 2D Hubbard model
and QMC results exhibit quantitative agreement. In fact, the duality reveals that the
Ising variables of the FKM are exactly the constraints of the slave-spin
representation. Considerations regarding the relevance of the latter lead to
the generalized phase diagram of the unconstrained \cite{PhysRevB.96.205104}
slave-spin Hubbard model shown in Fig.~\ref{fig:grandphasediagram} and discussed in Sec.~\ref{sec:model}.

Here, we provide a significantly more detailed account of the different
representations, the relevant symmetries, and the relation to
the slave-spin theory of the Hubbard model. The distinct regimes of the
phase diagram are characterized with a focus on $T>T_Q$.
In particular, we study the temperature dependence of the
compressibility, the conductivity, and the optical conductivity. The
appearance of Hubbard physics at $T<T_Q$, but also the clearly non-Hubbard
physics at $T>T_Q$, are illustrated. The results are followed by a discussion of
the interpretation in terms of slave-spin theory and OMs,
especially the differences between the present realization and mean-field theories
or exactly solvable models. 

The rest of this paper is organized as follows. In Sec.~\ref{sec:model} we
discuss the Hamiltonians. The QMC method is outlined in
Sec.~\ref{sec:method}. Numerical results are reported in
Sec.~\ref{sec:results}, followed by a discussion in
Sec.~\ref{sec:discussion}. Finally, we conclude in
Sec.~\ref{sec:conclusions}.

\section{Model, dualities, symmetries}\label{sec:model}

\subsection{Hamiltonians}\label{sec:hamiltonians}

The square-lattice FKM introduced in Ref.~\RefI is defined by the Hamiltonian
\begin{align}\label{eq:HcQ}\nonumber
  \hat{H}^{cQ} 
  = 
  &-t \sum_{\las ij\ras,\sigma} \left( \oc^\dag_{i\sigma} \oc^\nag_{j\sigma} + \text{H.c.} \right)    
    \\
    &- U \sum_{i} \left(\on_{i\UP}-\oh\right) \left(\on_{i\DO}-\oh\right)  \oQ_i\,.
\end{align}
The first term describes the hopping of electrons with spin $\sigma$
between nearest-neighbor lattice sites $i$ and $j$ with amplitude $t$. The electrons
couple to Ising degrees of freedom $\oQ_i=\pm 1$ located at the sites with
strength $U$. The density operator is defined as
$\on^{}_{i\sigma}=\oc^\dag_{i\sigma}\oc^{}_{i\sigma}$, the chemical potential
$\mu=0$ for the case of half-filling ($\las \on_i\ras=\las \on_{i\UP} +
\on_{i\DO}\ras =1$) considered here.

Writing the interaction as $U \sum_{i} \oQ_i \prod_\alpha
(\on_{i\alpha}-\oh)$ with a flavor index $\alpha$, Eq.~(\ref{eq:HcQ}) may be
regarded as a generalization of the spinless FKM with one flavor of itinerant fermions 
that can be solved exactly in infinite dimensions \cite{RevModPhys.75.1333}. 
In fact, the original FKM of Ref.~\cite{falicov1969simple} was formulated
with itinerant and localized SU(2) fermions, but does not contain a
three-body interaction as in Eq.~(\ref{eq:HcQ}). The latter seems to 
preclude an exact solution but, as discussed in Sec.~\ref{sec:method},
Hamiltonian~(\ref{eq:HcQ}) with $U>0$ is amenable to sign-free QMC simulations. 

The connection to other FKMs becomes complete if the local Ising variables
$\oQ_i$ are expressed in terms of the occupation numbers of localized spinless fermions via 
\begin{equation}\label{eq:Qtol}
\oQ_i = 1 - 2 \on^{l}_i = 1 - 2 \hat{l}^\dag_i \hat{l}^\nag_i \,.
\end{equation}
This yields an onsite interaction $\sim U \on_{i\UP} \on_{i\DO} \on^l_{i}$.

Whereas we consider a half-filled band for the $c$ fermions, the number
of $l$ fermions is determined by their interaction with the $c$ fermions and
depends on $U$ and temperature. It is directly
related to the magnetization of the $\oQ_i$ variables. 
 A ferromagnetic state with $\las \oQ_i\ras=1$
($\las \oQ_i\ras=-1$) corresponds to $\las \on^l_i\ras=0$  ($\las \on^l_i\ras
=1$), whereas a paramagnetic state with $\las \oQ_i\ras=0$ implies $\las \on^l_i \ras=0.5$.

The representations based on either the fermion occupation numbers $\on^\text{l}_i$ or
the Ising variables $\oQ_i$ (both of which are conserved by the respective
Hamiltonian) should yield identical results for $c$-fermion
properties. However, only the fermionic representation reveals the nontrivial
spectral properties of the $l$ fermions in the spinless FKM
\cite{brandt1992thef,PhysRevB.59.2642,PhysRevB.71.115111} via
$\las \hat{l}^\dag_i(\tau)\,\hat{l}_i\ras$ \footnote{We thank
G. Czycholl for pointing this out to us.}.

\begin{figure}[t]
\includegraphics[width=0.45\textwidth]{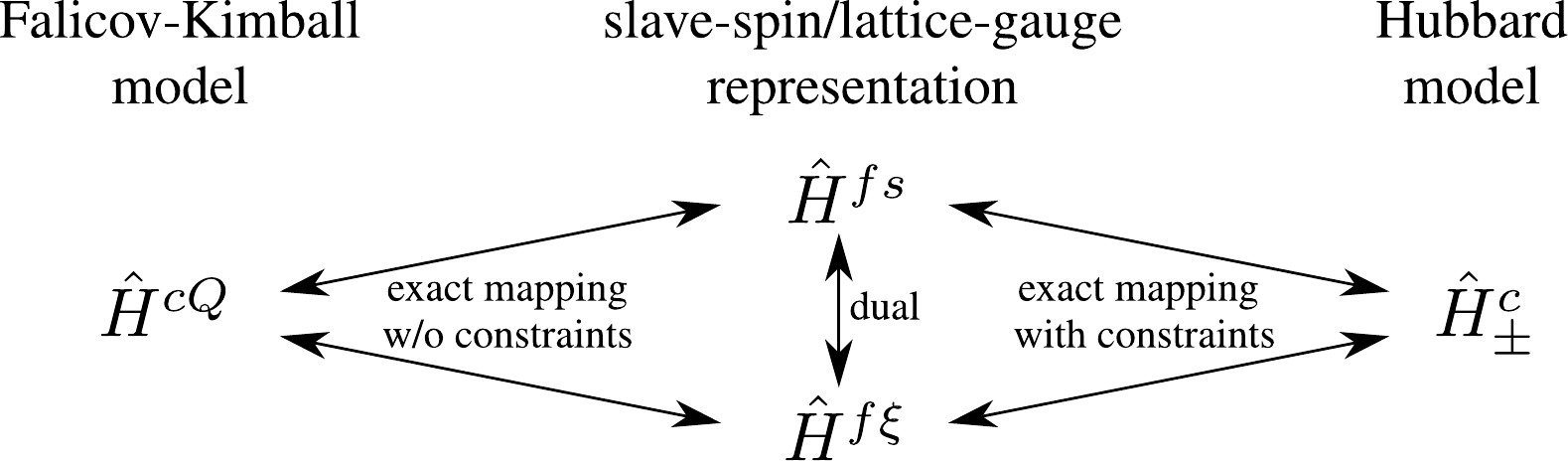}
\caption{\label{fig:relations}
The relations between the Hamiltonians of Sec.~\ref{sec:model}. The
2D FKM~(\ref{eq:HcQ}) is dual to an unconstrained slave-spin or lattice gauge
theory that can be written either in terms of site Ising spins
[Eq.~(\ref{eq:Hfs})] or bond Ising spins [Eq.~(\ref{eq:Hftau})]. If Gauss's
law, defined by the local constraints~(\ref{eq:constraint2}), holds,
Eqs.~(\ref{eq:Hfs}) and~(\ref{eq:Hftau}) become constrained gauge theories
or, equivalently, exact slave-spin representations of the Hubbard
model~(\ref{eq:hubbard}).
}
\end{figure}

\begin{figure}[t]
  \includegraphics[width=0.325\textwidth]{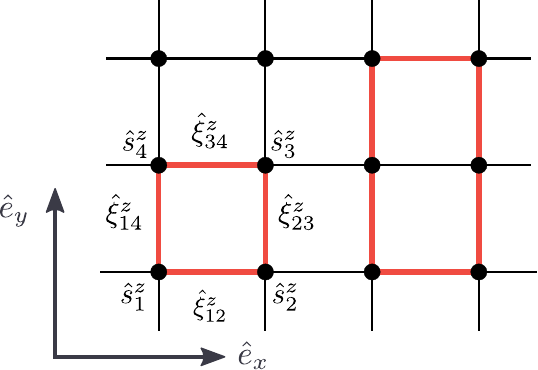}  
  \caption{\label{fig:isingspins} Relation between the
    bond Ising variables in Eq.~(\ref{eq:Hftau}) and the site Ising
    variables in Eq.~(\ref{eq:Hfs}). For any closed loop (indicated by thick
    red lines), $\prod_\square
    \hat{\xi}^z_{ij} = +1$ independent of the slave-spin configuration.
  }
\end{figure}

As illustrated in Fig.~\ref{fig:relations}, the FKM~(\ref{eq:HcQ}) is related by an exact duality to a model of fermions
coupled to Ising spins in a transverse field, which itself has two equivalent
or dual representations. The first of these dual Hamiltonians is obtained by
making the operator replacements
\begin{equation}\label{eq:ctofs}
  \oc^\dag_{i\sigma} \mapsto \of^\dag_{i\sigma}\os^z_i\,,\quad
  \oc^{}_{i\sigma} \mapsto \of^{}_{i\sigma}\os^z_i\,,
\end{equation}
and
\begin{equation}\label{eq:Qtosf}
  \oQ_i = \os^x_i (-1)^{\sum_\sigma \of^\dag_{i\sigma} \of^{}_{i\sigma}}
        = \os^x_i (-1)^{\sum_\sigma \oc^\dag_{i\sigma} \oc^{}_{i\sigma}}\,,
\end{equation}
where $\of^\dag_{i\sigma} \of^{}_{i\sigma}= \oc^\dag_{i\sigma} \oc^{}_{i\sigma}$ follows from Eq.~(\ref{eq:ctofs}).
Here, $\of^\dag_{i\sigma}$ creates a fermion with spin $\sigma$ at site $i$,
whereas $\os^z_i$ and $\os^x_i$ are represented by Pauli matrices acting on
an Ising spin at site $i$. Using the identity
\begin{equation}
  (-1)^{\sum_\sigma \of^\dag_{i\sigma} \of^{}_{i\sigma}} = \prod_\sigma (2\on_{i\sigma}-1)
\end{equation}
we can rewrite Eq.~(\ref{eq:HcQ}) as
\begin{equation}\label{eq:Hfs}
  \hat{H}^{fs} 
  = 
  -t \sum_{\las ij\ras,\sigma} \left( \of^\dag_{i\sigma} \of^\nag_{j\sigma} \os^z_i
  \os^z_j + \text{H.c.} \right)    
    - \frac{U}{4} \sum_{i} \os^x_i\,.
\end{equation}
Equation~(\ref{eq:Hfs}) has a dual representation in terms of fermions and
{\em bond} Ising variables. The mapping between site Ising variables $\os^\alpha_i$ and bond Ising
variables $\oxi^\alpha_{ij}$ is illustrated in Fig.~\ref{fig:isingspins}. For nearest-neighbor
sites $i$ and $j$, 
\begin{equation}\label{eq:bondtosite}
\os_i^z \os^z_j\mapsto \oxi^z_{ij}\,.
\end{equation}
Because a spin flip on a single site $i$ under the action of $\os^x_i$
affects all four bond variables, the dual representation involves a so-called
star operator, 
\begin{equation}
\os^x_i \mapsto 
\hat{\xi}^x_{i,i+x} \hat{\xi}^x_{i,i-x} \hat{\xi}^x_{i,i+y}  \hat{\xi}^x_{i,i-y}\,,
\end{equation}
where $i\pm\alpha$ is a compact notation for the site at $\bm{r}_i \pm \hat{e}_\alpha$.
These steps lead to the Hamiltonian
\begin{align}\label{eq:Hftau}
  \hat{H}^{f\xi}
  = 
  &-t \sum_{\las ij\ras,\sigma}
  \left(\of^\dag_{i\sigma}\of^\nag_{j\sigma} +
  \text{H.c.}\right)\,\hat{\xi}^z_{ij} \\\nonumber
  &-
  \frac{U}{4}\sum_i
    \hat{\xi}^x_{i,i+x} \hat{\xi}^x_{i,i-x}
    \hat{\xi}^x_{i,i+y}  \hat{\xi}^x_{i,i-y} 
\,.
\end{align}
Hamiltonian~(\ref{eq:Hftau}) with bond Ising variables $\oxi^\alpha_{ij}$
takes the form familiar from Ising lattice gauge theories coupled to matter.  However,
there are no additional constraints (also known as Gauss's law) in the
context of the FKM. Such constraints arise in slave-spin representations (see
Sec.~\ref{sec:slavespin}) and their role for the physics observed will be
addressed in detail in Sec.~\ref{sec:discussion}. 

The origin of the lattice gauge theory~(\ref{eq:Hftau}) in a slave-spin
representation implies the absence of $\pi$-flux configurations of the bond Ising variables.
As illustrated in Fig.~\ref{fig:isingspins}, on any closed loop, the sign
of a site Ising variable $\os^z_i$ affects the sign of an even number of bond
variables $\oxi^z_{ij}$ in the loop [cf. the star operator in
Eq.~(\ref{eq:Hftau})]. Therefore, the product of the $\oxi^z_{ij}$
over the loop will always equal $+1$, corresponding to the absence of a
$\pi$ flux. Because of the link between site and bond variables,
Eq.~(\ref{eq:bondtosite}), the latter cannot be flipped individually.
The absence of $\pi$ fluxes was accounted for in our simulations based on
Eq.~(\ref{eq:Hftau}) by a proper choice of update rules, see
Sec.~\ref{sec:method}. 

\subsection{Relation to the slave-spin Hubbard model}\label{sec:slavespin}

The classification of the unusual metallic phase observed at $T>T_Q$ and
$U\gtrsim 6$ (see Fig.~\ref{fig:phasediagram}) as a fractionalized metal is based
on close conceptual relations with slave-spin theories and OMs. Most importantly,
Eq.~(\ref{eq:Hftau}) is
identical to the Ising (\ZII) slave-spin representation of the Hubbard
Hamiltonian (we allow both interaction signs)
\begin{align}\label{eq:hubbard}
\hat{H}^c_{\pm} 
= 
&-t \sum_{\las ij\ras,\sigma} ( \oc^\dag_{i\sigma} \oc^\nag_{j\sigma} + \text{H.c.})
% - \mu\sum_i \on_i
%\
 \pm U\sum_i (\on_{i\UP}-\oh)(\on_{i\DO}-\oh) \,;
\end{align}
taking $U>0$, $\hat{H}^c_{+}$ ($\hat{H}^c_{-}$) corresponds to
repulsive (attractive) interactions. This identification relies on the highly
symmetric form of interactions in the FKM~(\ref{eq:HcQ}) with only a single
interaction parameter $U$, in contrast to more general multi-component FKMs
with independent interaction matrix elements and onsite energies
\cite{RevModPhys.75.1333,PhysRevB.96.205104}.

In complete analogy with Eq.~(\ref{eq:ctofs}), the idea of the slave-spin
representation is to replace the original fermionic operators by {\em auxiliary
fermions} $f$ and {\em slave Ising spins $s$}. The orientation of the slave
spin at a given site is supposed to distinguish between doubly occupied and
empty configurations on the one hand, and singly-occupied (local moment)
configurations on the other hand. One of two possible choices is
$\ket{\UP}^s_i \leftrightarrow n_i=0,2$ (even parity) and
$\ket{\DO}^s_i\leftrightarrow n_i=1$ (odd parity), where $n_i$ is the
eigenvalue of 
\begin{equation}\label{eq:occupationnumber}
\on_{i} 
=
\sum_\sigma \oc^\dag_{i\sigma} \oc^\nag_{i\sigma} 
=
\sum_\sigma \of^{\dag}_{i\sigma} \os^z_i \of^{\nag}_{i\sigma} \os^z_i
=
\sum_\sigma \of^{\dag}_{i\sigma} \of^{\nag}_{i\sigma}\,.
\end{equation}
Because the hopping (interaction) term in Eq.~(\ref{eq:hubbard}) preserves
(changes) the occupation number at the involved lattice sites, the slave-spin
transformation is usually chosen as $\oc^{(\dag)}_{i\sigma} \mapsto
\of^{(\dag)}_{i\sigma} \os^x_i$. In the present work, we prefer to use the
same form of the Hamiltonian throughout, and instead change the Ising
quantization basis. Hence, we use the replacement
\begin{equation}\label{eq:slave-spin1}
  \oc^{(\dag)}_{i\sigma} \mapsto \of^{(\dag)}_{i\sigma} \os^z_i\,,
\end{equation}
which immediately yields the same hopping term as in Eq.~(\ref{eq:Hfs}). To
make contact with the usual slave-spin picture, we can consider specifying
the slave-spin orientation relative to the $x$ axis, thereby obtaining a
diagonal Hubbard term. However, for the remainder of the paper, we exploit
the freedom of switching back to the $z$ basis. This makes the hopping term
diagonal in the slave spins and turns the Hubbard term into a transverse
field, which facilitates our auxiliary-field QMC
simulations. Moreover, the Hamiltonian (but not the physics) becomes
identical to that considered in Refs.~\cite{PhysRevX.6.041049,gazit2017emergent}.

In order to replace the Hubbard interaction by the simpler transverse-field term,
\begin{equation}\label{eq:docctotransversefield}
U\sum_i \left(\on_{i\UP}-\oh\right)\left(\on_{i\DO}-\oh\right) \mapsto \frac{U}{4} \sum_i \os^x_i\,,
\end{equation}
we have to assume that the slave spin at a site $i$ faithfully describes the
parity of $n_i$. Because auxiliary fermions and slave spins are formally independent,
$[\of^{(\dag)}_{i\sigma},\os^\alpha_j]=0$, the local Hilbert space in the
slave-spin representation is
\begin{equation*}
 \mathcal{H}^f_i\otimes\mathcal{H}^s_i =
 \{\ket{0}_i,\ket{\UP}_i,\ket{\DO}_i,\ket{\UP\DO}_i\}_f\otimes\{\ket{\UP}_i,\ket{\DO}_i\}_s
\end{equation*}
and has a dimension twice as large as that of the original Hubbard model where
$\mathcal{H}^c_i=\{\ket{0}_i,\ket{\UP}_i,\ket{\DO}_i,\ket{\UP\DO}_i\}_c$.
To select the four physical states and thereby achieve an exact
representation of the Hubbard model, the slave-spin Hamiltonian~(\ref{eq:Hfs})
has to be supplemented with a local constraint on the corresponding eigenvalues.
The latter can be formulated in several equivalent ways, including
\begin{equation}\label{eq:constraint1}
  \os^x_i - (-1)^{\on_i} \stackrel{!}{=} 0\,.
\end{equation}
An alternative form, obtained by squaring Eq.~(\ref{eq:constraint1}), reveals
a direct connection to the FKM:
\begin{equation}\label{eq:constraint2}
  \oQ_i = \os^x_i (-1)^{\on_i}\stackrel{!}{=} +1\,.
\end{equation}
The choice $+1$ gives the above identification of $\ket{\UP}^s_i$ with $n_i=0,2$,
whereas the physically equivalent choice $-1$ amounts to $\ket{\DO}^s_i$ representing $n_i=0,2$.
In the FKM representation~(\ref{eq:HcQ}), the locally conserved $\oQ_i$ may
be replaced by classical variables $Q_i$. Their operator nature arises only
after fractionalization of the $c$ fermions and is in particular reflected in the nontrivial
commutation relations in Eq.~(\ref{eq:QlocalIsingtransformation}) below.

The necessity of local constraints is a fundamental difference between
the slave-spin representation of the Hubbard model and the corresponding
transformation for the FKM that leads from Eq.~(\ref{eq:HcQ}) to Eq.~(\ref{eq:Hfs}).
For the FKM, the local Hilbert space dimension remains unchanged, so that the
transformation amounts to a mere relabeling of the states. Explicitly, for the FKM~(\ref{eq:HcQ}),
\begin{equation*}
  \mathcal{H}^c_i\otimes\mathcal{H}^Q_i =
  \{\ket{0}_i,\ket{\UP}_i,\ket{\DO}_i,\ket{\UP\DO}_i\}_c\otimes\{\ket{+1}_i,\ket{-1}_i\}_Q\,,
\end{equation*}
whereas for the fermion-spin model~(\ref{eq:Hfs}) 
\begin{equation*}
 \mathcal{H}^f_i\otimes\mathcal{H}^s_i =
 \{\ket{0}_i,\ket{\UP}_i,\ket{\DO}_i,\ket{\UP\DO}_i\}_f\otimes\{\ket{\UP}_i,\ket{\DO}_i\}_s\,.
\end{equation*}

The unconstrained slave-spin representation of the Hubbard model is
identical to the slave-spin representation~(\ref{eq:Hfs}) of the FKM (where no constraints
arise).  The addition of
the constraints~(\ref{eq:constraint2}) promotes Hamiltonian~(\ref{eq:Hfs})
from an unconstrained gauge theory to a proper, constrained lattice gauge
theory or, equivalently, an exact slave-spin representation of Eq.~(\ref{eq:hubbard}).

The fact that Eq.~(\ref{eq:constraint2}) resembles Eq.~(\ref{eq:Qtosf}) for the FKM allows us to identify the Ising
variables $\oQ_i$ of the latter with the constraints of the slave-spin representation of the Hubbard model. Their role
in ensuring a one-to-one mapping between Eqs.~(\ref{eq:hubbard})
and~(\ref{eq:Hfs}) becomes particularly clear from the observation that
setting $\oQ_i=1$ ($\oQ_i=-1$) in Eq.~(\ref{eq:HcQ}) directly reduces the FKM to the
attractive (repulsive) Hubbard model. 
In the context of the FKM, the constraints are spontaneously generated at the
transition temperature $T_Q$, see Fig.~\ref{fig:phasediagram}. Therefore, at $T=0$, the FKM is equivalent to
the Hubbard model or, alternatively, a constrained slave-spin gauge
theory. In Sec.~\ref{sec:results:hubbard}, we will demonstrate numerically that in the ferromagnetic
phase at $T<T_Q$, the FKM does indeed yield results that are in quantitative agreement
with those for the Hubbard model.

The fully polarized ferromagnetic state of the FKM is only
one of several nontrivial limits in which an exact slave-spin or constrained
lattice gauge representation emerges without explicitly imposing the constraints.
In fact, in all the shaded regions in Fig.~\ref{fig:grandphasediagram}, the
constraints are either irrelevant or dynamically generated and hence not
necessary for a faithful representation of the Hubbard model.
For example, for $U=0$, Eq.~(\ref{eq:HcQ}) immediately reveals that the
$\oQ_i$ play no role. As pointed out in Ref.~\cite{PhysRevX.6.041049}, all
$\hat{Q}_i$ sectors are degenerate in this limit.
This is not only true for the ground state, as previously noted in Ref.~\cite{PhysRevB.81.155118}, but
at any temperature. Numerical evidence will be presented in
Sec.~\ref{sec:results:hubbard}.

Much more remarkably, the constraints are also irrelevant for
any $U$ and $T$ in the limit of large spatial dimensions, $\text{D}=\infty$
\cite{schiro2011quantum}.  Again, this
becomes particularly transparent from the perspective of the FKM.
A weak-coupling expansion of the partition function in powers of $\oH_1 =
U\sum_i \left(\on_{i\UP}-\oh\right) \left(\on_{i\DO}-\oh\right) \oQ_i$ takes the form 
\begin{equation}\label{eq:expansion}
\frac{Z}{Z_0} 
= 
\sum_{n=0}^\infty\frac{(-1)^n}{n!} 
\left(\prod_{p=1}^n \int_0^\beta d\tau_p\right)
\las T \oH_1(\tau_1) \cdots \oH_1(\tau_n) \ras \,.
\end{equation}
The expectation value can be Wick-decomposed into all possible products of
real-space single-particle propagators. However, for $\text{D}=\infty$,
only the Hartree contributions are nonzero because nonlocal propagators
vanish as $1/\sqrt{\text{D}}$ or faster
\cite{muller1989correlated,khurana1990electrical}. This leads to an effective
Anderson impurity problem with a single interacting site $i_0$ to be solved
self-consistently. At half-filling, only even powers $n=2m$ occur in the
expansion~(\ref{eq:expansion})
due to particle-hole symmetry. In combination with $i_p=i_0$ for all $p$, the
constraints only enter in the form $(\oQ_{i_0})^{2m} \equiv 1$ and
the expansion reduces to that of the Hubbard model.

Because the constraints are irrelevant for $\text{D}=\infty$, the phase diagram of
Fig.~\ref{fig:grandphasediagram} for the {\em unconstrained} slave-spin
theory of the Hubbard model [Eq.~(\ref{eq:Hfs})] exhibits exactly the same
Mott metal-insulator transition as found by dynamical mean-field theory (DMFT)
for the repulsive Hubbard model~(\ref{eq:hubbard})
\cite{PhysRevLett.83.3498,PhysRevLett.101.186403,PhysRevB.91.245130,PhysRevLett.114.246402}.
In the slave-spin representation, the $T=0$ transition at $U_c>0$ is associated
with a phase transition of the slave spins \cite{PhysRevB.91.245130}. At $T>0$, a line of
first-order transitions terminates at a critical endpoint at $T_c$, above
which a metal-insulator (or FL to bad metal) crossover remains
\cite{PhysRevLett.83.3498,PhysRevLett.101.186403,PhysRevB.91.245130,PhysRevLett.114.246402}.

\subsection{Symmetries and their consequences}\label{sec:model:symmetries}

A look at the symmetries of the problem will prove useful for understanding
the numerical results. First, the FKM~(\ref{eq:HcQ}) on
the bipartite square lattice is invariant under the particle-hole
transformation (acting on $\sigma=\UP,\DO$)
\begin{equation}
  \hat{P}^{-1}\,\oc^\dag_{i\sigma}\,\hat{P}^{} = \eta_i \oc^{}_{i\sigma}\,,\quad
  \hat{P}^{-1}\,\oc^{}_{i\sigma}\,\hat{P}^{} = \eta_i \oc^{\dag}_{i\sigma}\,,
\end{equation}
with the sublattice dependent phase factor
\begin{equation}
  \eta_i = e^{\rmi(\pi,\pi)\cdot\vec{r}_i} = \pm 1\,.
\end{equation}

The SO(4)$=$SU(2)$\times$SU(2)/\ZII symmetry of the Hubbard model, composed
of SU(2) charge and spin symmetries \cite{yang1990so}, is  present in
the FKM as a subgroup of a larger O(4) symmetry. This
follows from the fact that under the partial particle-hole or Shiba
transformation \cite{essler2005one}, defined by
\begin{equation}\label{eq:phtrafo}
  \hat{P}^{-1}_{{\sigma}}\,\oc^\dag_{i\sigma'}\,\hat{P}^{}_{{\sigma}}
  = 
  \delta_{\sigma{\sigma}'}\,e^{\rmi(\pi,\pi)\cdot\vec{r}_i} \oc^{}_{i\sigma}
  + 
  (1 -  \delta_{\sigma{\sigma}'})\oc^\dag_{i\sigma}\,,
\end{equation}
the hopping term remains unchanged but $U\to-U$ in the interaction term. The
transformation hence connects repulsive and attractive Hubbard models, involving an interchange
of charge and spin-$z$ operators \cite{yang1990so}. In the FKM,
we can compensate the sign change of $U$ by the global transformation
$\oQ_i\mapsto -\oQ_i$. This constitutes an additional \ZII symmetry (often
referred to as parity or particle-hole symmetry) absent in
the Hubbard model that restores the global $\text{SO}(4)\times\text{Z}_2=\text{O}(4)$ symmetry
of free fermions. The latter implies, in particular, identical
spin and charge correlation functions. Even though we explicitly choose an attractive
interaction in Eq.~(\ref{eq:HcQ}) to avoid a QMC sign problem, the physics in the interesting
O(4) symmetric regime at $T>T_Q$ is independent of the interaction sign.

The additional \ZII symmetry is broken by the Hubbard term in
Eq.~(\ref{eq:hubbard}). However, the Hubbard interaction in the 
slave-spin representation, $\hat{H}_U \sim \sum_i \os^x_i$, is invariant
under an equivalent Shiba transformation for the $f$ fermions, leading
to an O(4) symmetry at the level of Hamiltonian~(\ref{eq:Hfs}). For the
Hubbard model, the constraints [Eq.~(\ref{eq:Qtosf}) or
Eq.~(\ref{eq:constraint2})] change sign under $\hat{P}^{}_{\sigma}$,
\begin{equation}\label{eq:QunderPH}
  \hat{P}^{-1}_{{\sigma}}\,\oQ^{}_{i}\,\hat{P}^{}_{{\sigma}} = -\oQ^{}_i\,,
\end{equation}
since, for example, 
$\hat{P}^{-1}_{\UP}(-1)^{\sum_\sigma\on_{i\sigma}}\hat{P}^{}_{\UP}=(-1)^{1-\on_{i\UP}+\on_{i\DO}}\,.$
Therefore, the constrained slave-spin representation has the same SO(4)
symmetry as the original Hubbard Hamiltonian~(\ref{eq:hubbard}). 

The additional \ZII symmetry of the FKM is spontaneously broken at $T<T_Q$. In the Hubbard
regime, we hence have the same SO(4) symmetry as for the Hubbard
model. Similarly, the \ZII symmetry can be broken by a magnetic field  $-h_Q
\sum_i \oQ_i$ or, as in Sec.~\ref{sec:results}, by a Hubbard
term. By the Mermin-Wagner theorem, spontaneous symmetry breaking in the
fermionic sector only takes place at $T=0$, where the FKM has the same ground state as the Hubbard model:
long-range AFM order for repulsive interactions ($\las\oQ_i\ras=-1$) and combined CDW/SC
order with an SO(3) order parameter for attractive interactions ($\las \oQ_i\ras=+1$).
In contrast, long-range order occurs also at $T>0$ in a recent DMFT study of
Eq.~(\ref{eq:HcQ}) \cite{tran2018fractionalization}.

As suggested by the connection to lattice gauge theories, the slave-spin
representation introduces a local \ZII symmetry. We can change the sign of
both $\of^{(\dag)}_{i\sigma}$ and $\os^z_i$ while leaving
$\oc^{(\dag)}_{i\sigma}=\of^{(\dag)}_{i\sigma} \os^z_i$ unchanged. The generators
of this transformation are the constraints $\oQ_i$ with the
property $\hat{Q}_i=\hat{Q}^{-1}_i$ since $\hat{Q}_i^2=1$. The local symmetry
is reflected in
\begin{equation}\label{eq:hqcommute}
  [\hat{H}^{fs},\hat{Q}_i] = [\hat{H}^{cQ},\hat{Q}_i] = 0\,.
\end{equation}
The operators in Eq.~(\ref{eq:Hfs}) transform as
\begin{align}\label{eq:QlocalIsingtransformation}
  \nonumber
  \of^{(\dag)}_{i\sigma}
  &\mapsto
  \hat{Q}_i \of^{(\dag)}_{i\sigma} \hat{Q}_i = - \of^{(\dag)}_{i\sigma}\,,\quad
  \\    
  \nonumber
  \os^z_i
  &\mapsto 
  \hat{Q}_i \,\,\os^z_i \,\,\hat{Q}_i = -\os^z_i\,.
  \\    
  \os^x_i
  &\mapsto 
  \hat{Q}_i \,\,\os^x_i \,\,\hat{Q}_i = +\os^z_i\,.
\end{align}
The $f$ fermions and the slave spins
$\os^z_i$  carry a \ZII charge, whereas $\of^{(\dag)}_{i\sigma}
\os^z_i\equiv \oc^{(\dag)}_{i\sigma}$ and $\os^x_i$ are neutral. 
Although invariance under local transformations is in the present setting
related to an actual symmetry, the same operators
appear in slave-spin theories and true gauge theories where the local
transformations are associated with gauge invariance (choice of
representation). Therefore, we will not strictly distinguish between \ZII and
gauge charges or \ZII and gauge transformations.

According to Elitzur's theorem \cite{PhysRevD.12.3978,RevModPhys.51.659},
local symmetries cannot be spontaneously broken in finite dimensions
at $T>0$. At $T=0$, symmetry breaking requires a
macroscopic degeneracy of the ground state. This is realized at the critical
point for the 1D Ising model \cite{Subrahmanyam_1988}. However, in finite
dimensions, we generally expect the third law of thermodynamics to hold and
hence Elitzur's theorem to apply at any temperature. 

The local symmetry imposes significant restrictions on correlation functions. 
Equation~(\ref{eq:hqcommute}) yields
\begin{align}\nonumber\label{eq:qfq}
  \hat{Q}_i \of^{(\dag)}_j  \hat{Q}_i &= (1-\delta_{ij}) \of^{(\dag)}_j\,, \\
  \hat{Q}_i \os^{z}_j  \hat{Q}_i &= (1-\delta_{ij}) \os^z_j\,,
\end{align}
as well as $\hat{Q}_i \os^{x}_j  \hat{Q}_i = \os^x_j$. Using
Eqs.~(\ref{eq:constraint2}) and~(\ref{eq:qfq}), $\hat{Q}_i^2=1$,
 as well as the cyclic invariance of the trace, leads to
\begin{align}\nonumber\label{eq:fandtaulocal}
  \las \of^\dag_{i\sigma}(\tau) \of^\nag_{j\sigma}(0) \ras 
  &=
  \delta_{ij}  \las \of^\dag_{i\sigma}(\tau) \of^\nag_{i\sigma}(0) \ras\,,\\
  \las \os^z_{i\sigma}(\tau) \os^z_{j\sigma}(0) \ras 
  &=
  \delta_{ij}  \las \os^z_{i}(\tau) \os^z_{i}(0) \ras\,.
\end{align}
Hence, correlation functions of gauge-dependent (charged) operators are
entirely local in space. The key role of the local \ZII symmetry in our model
and in slave-spin theories will be discussed in Sec.~\ref{sec:gauge-symmetry}.

Finally, in contrast to Eq.~(\ref{eq:Hfs}), the FKM~(\ref{eq:HcQ}) 
contains only neutral operators and has a global Ising symmetry. A local U(1) symmetry (containing the \ZII
subgroup) emerges if the Ising variables are replaced by local
fermions according to Eq.~(\ref{eq:Qtol}), reflecting invariance under the
transformation
$\hat{l}^\dag_i \mapsto e^{\rmi\varphi } \hat{l}^\dag_i$,
$\hat{l}^\nag_i \mapsto e^{-\rmi\varphi } \hat{l}^\nag_i$  \cite{RevModPhys.75.1333}.

\section{Method}\label{sec:method}

QMC simulations were carried out using the implementation of the 
auxiliary-field method from the Algorithms for
Lattice Fermions (ALF) library \cite{ALF17}. This algorithm
for correlated fermions goes back to the
work of Blankenbecler \etal~\cite{Blankenbecler81} and is based on a
stochastic evaluation of the fermionic path integral with the help of
discrete auxiliary fields. For a review see Ref.~\cite{Assaad08_rev}.

The simulations were done in the lattice-gauge representation~(\ref{eq:Hftau}). The partition function can be written as a
Euclidean path integral over configurations $\bm{\xi}=\{{\xi}^z_{ij,\tau}\}$,
\begin{equation}
  Z = \tr\, e^{-\beta (\hat{H}-\mu \hat{N})} = \int \mathcal{D}[\bm{\xi}]\,e^{-S[\bm{\xi}]}\,.
\end{equation}
As usual, imaginary time was discretized with a Trotter timestep
$\Delta\tau=\beta/L$. The latter was typically fixed to $U\Delta\tau=0.1$,
but much smaller values were used for simulations at high temperatures in
order to have a sufficient resolution in imaginary time.

The configuration weight can be written as $e^{-S} = e^{-S_0}\det[1 +
B(\bm{\xi})]$. Here, the action $S_0$ describes the spin dynamics due to the 
transverse field, whereas $B$ is a product over time slices of exponentials
of the hopping part that contains the fermion-spin coupling. Because $S$ is
real, there is no sign problem for $U>0$.

An important distinction between the present problem and QMC simulations of
Hubbard-Stratonovitch fields or fermions coupled to bond Ising spins is the
fact that the bond spins $\hat{\xi}^z_{ij}$ are composed of two slave spins
$\os^z_i$ and $\os^z_j$ at sites $i$ and $j$. Therefore, the
minimal update consistent with the physics of the problem corresponds to
flipping not one but all four bond spins connected to a given site. In particular,
this ensures the absence of $\pi$-flux configurations. Since the bond spins
determine the site spins only up to an overall sign, we stored a reference
value $s^z_{1}$ for each configuration.

For the FKM~(\ref{eq:HcQ}), simulations in
the bond-spin basis lead to a perfect sampling of the Ising variables
$\oQ_i$ and hence a symmetric distribution with $\las\oQ_i\ras=0$. Because
symmetry breaking only occurs for $L\to\infty$, our data therefore generally
exhibit an O(4) symmetry unless the latter is broken explicitly.
The results for the original Hubbard model in the standard fermionic
representation were obtained using an SU(2) symmetric decomposition of the
interaction \cite{ALF17}.

Observables were measured using the single-particle Green function and Wick's
theorem~\cite{Assaad08_rev}. In addition to the gauge-invariant observables
defined in terms of the original fermions, we also measured correlation
functions of the slave spins $\os^z_i$. Dynamic correlation functions were
accessible in imaginary time and analytically continued to real
frequencies for selected parameters using the stochastic maximum entropy method 
of Ref.~\cite{Beach04a}.

We work in units where $\hbar$, $\kB$, and $t$ are equal to one. All results
are for square lattices with $L\times L$ sites and periodic boundary conditions.

\section{Results}\label{sec:results}

The presentation of our data is organized as follows. First, we
recapitulate selected results of Ref.~\RefI to establish the key features of
the phase diagram. Then, we demonstrate the
emergence of Hubbard physics at $T<T_Q$,
before turning to the two different metallic regimes at $T>T_Q$.  Finally, we
discuss similarities and differences with respect to the Hubbard model.

\begin{figure}[t]
  \includegraphics[width=0.475\textwidth]{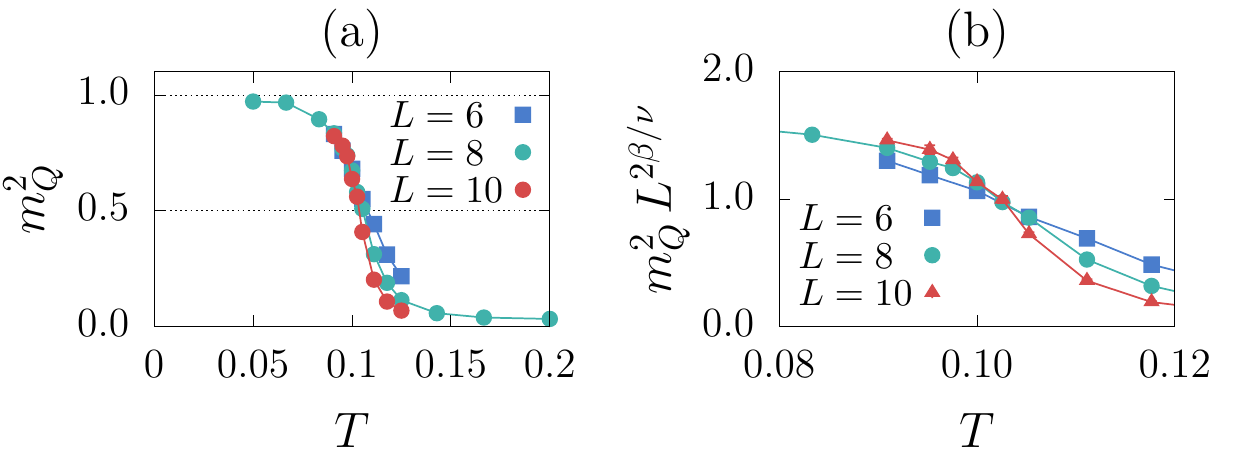}  
  \caption{\label{fig:constraints} Ferromagnetic phase transition in the FKM
    with $U=6$ from the squared magnetization $m^2_Q$ [Eq.~(\ref{eq:mq})]; $T_Q$ in
    Fig.~\ref{fig:phasediagram} was obtained from the condition
    $m^2_Q(T_{Q}){=}0.5$, as illustrated in (a). (b) 2D Ising
    critical behavior with exponents $\beta=1/8$ and $\nu=1$. Figure adapted
    from Ref.~\RefI.}
\end{figure}

\subsection{Structure of the phase diagram}\label{sec:struct-phase-diagr}

As illustrated in Fig.~\ref{fig:phasediagram}, the numerical
results support the existence of three distinct regimes in the $T$-$U$ phase
diagram of the FKM~(\ref{eq:HcQ}). This structure reflects the central role
of two different sets of Ising degrees of freedom in the problem---the constraints
$\oQ_i$ and the slave spins $\os^\alpha_i$---and their evolution with
temperature and the transverse field $h \sim U$.

The critical temperature $T_Q(U)$ separates a high-temperature phase with O(4)
symmetry and disordered Ising variables ($\las \oQ_i\ras=0$) from a
low-temperature phase with SO(4) symmetry and ferromagnetically ordered Ising
variables ($\las \oQ_i\ras\neq 0$). The phase transition originates
from a fermion-mediated exchange interaction $J \sum_{ij} \oQ_i \oQ_j$ that is
not forbidden by symmetry in Eq.~(\ref{eq:HcQ}) and will therefore be generated.
The ferromagnetic order of the $\oQ_i$ also lifts the macroscopic classical
degeneracy at $T=0$ and thereby avoids a violation of the third law of
thermodynamics \cite{PhysRevX.6.041049}. 

In our finite-size simulations, the transition  can be tracked by the square
of the magnetization per site,
\begin{equation}\label{eq:mq}
  m_Q^2 = \frac{1}{L^2} M^2_Q\,,\quad M^2_Q = \frac{1}{L^2}\sum_{ij}\las \oQ_i \oQ_j\ras\,.
\end{equation}
Figure~\ref{fig:constraints}(a) reveals that $m^2_Q\to 1$ ($m^2_Q\to 0$) at low
(high) temperatures. For simplicity,
the critical temperatures in Fig.~\ref{fig:phasediagram} were determined as
those where $m^2_Q$ equals $0.5$. The
comparison of $T_Q$ for different $L$ in Fig.~\ref{fig:constraints}(a) 
shows that finite-size effects are irrelevant for our purposes. Nevertheless,
we demonstrate in Fig.~\ref{fig:constraints}(b) that a similar $T_Q$ is obtained from the
crossing point in a proper finite-size scaling based on 2D Ising critical
exponents $\beta=1/8$ and $\nu=1$.

The shape of the phase boundary in Fig.~\ref{fig:constraints} is similar 
to that for the charge-density-wave (CDW) transition of the spinless FKM at
half-filling observed both for $\text{D}=2$ and $\text{D}=\infty$
\cite{RevModPhys.75.1333}. Exactly at $U=0$, the $\hat{Q}_i$ are decoupled
and do not order, so $T_Q=0$. A nonzero coupling generates an exchange interaction $J$ between the $\oQ_i$
and $T_Q\sim J$. For the spinless FKM, the exact solution in infinite dimensions
gives $J=U^2|\ln U|$ at weak coupling and $J\sim t^2/U$ at strong coupling
\cite{PhysRevLett.65.1663}, suggesting $T_Q\to0$  for $U\to\infty$ as in
Fig.~\ref{fig:phasediagram}.

\begin{figure}[t]
  \includegraphics[width=0.45\textwidth]{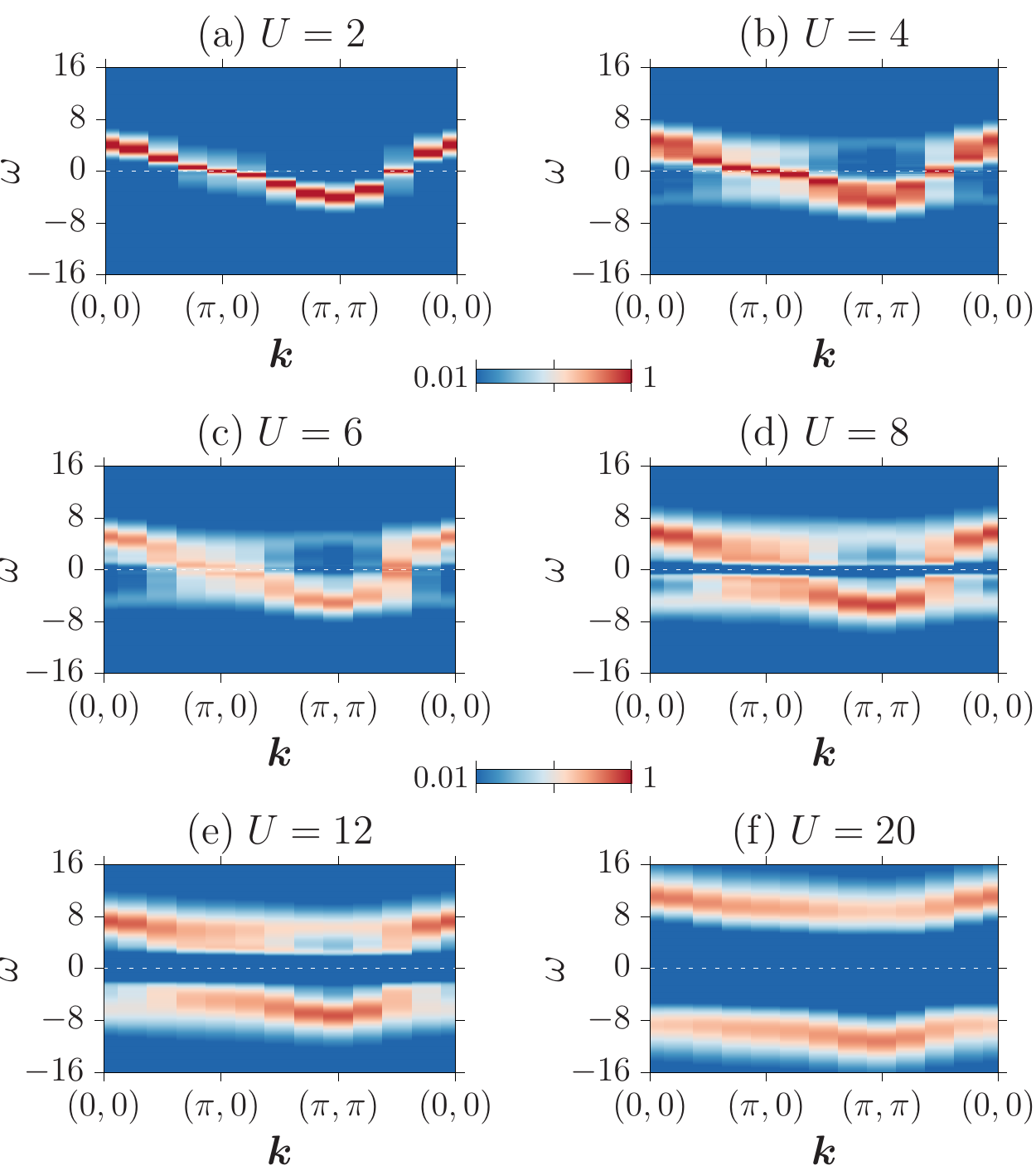} 
  \caption{\label{fig:akw-z2} Single-particle spectral function
    $A(\bk,\omega)$ [Eq.~(\ref{eqn:akw})] of the FKM as a function of $U$. Here, $L=8$,
    $T=1/6$.}
\end{figure}

\begin{figure}[t] 
\includegraphics[width=0.475\textwidth]{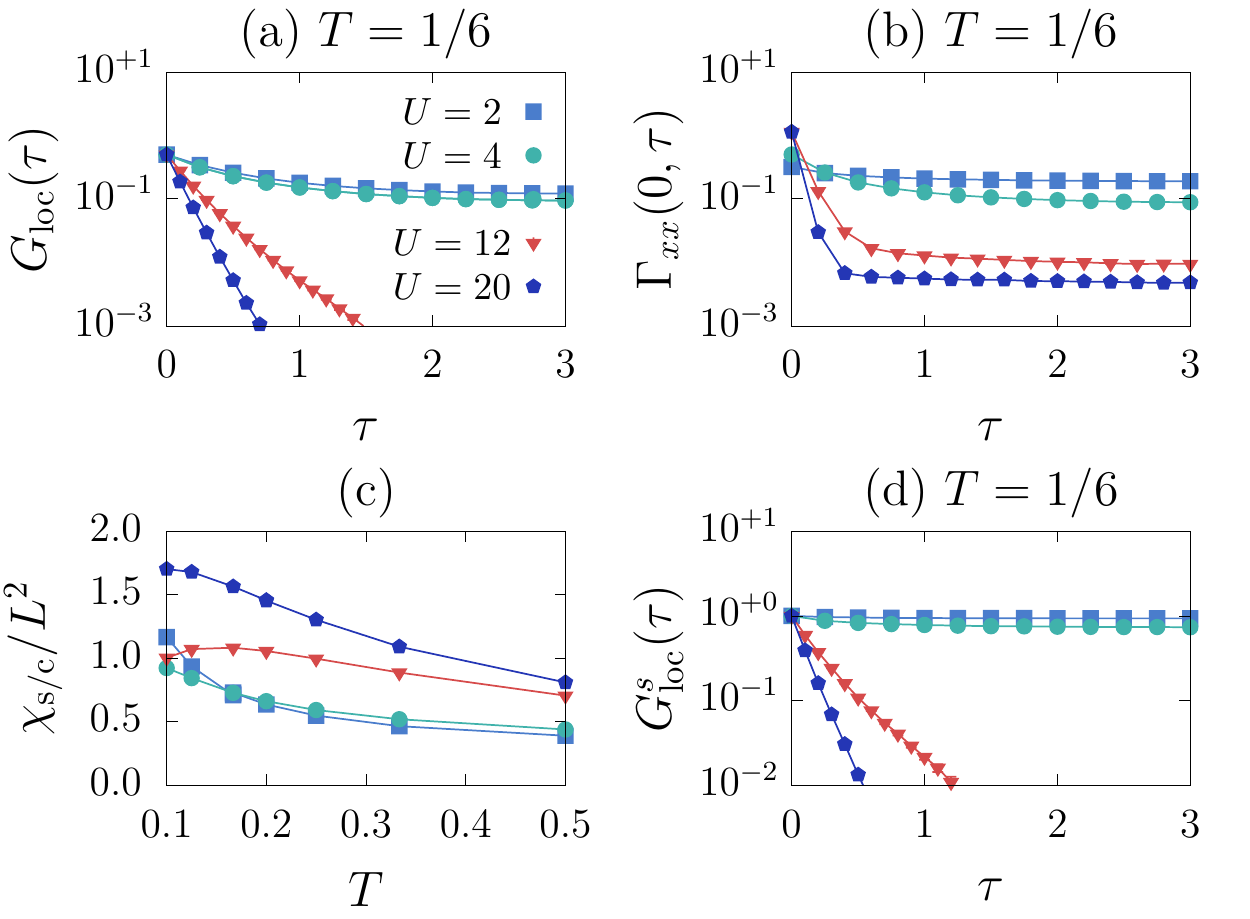}
 \caption{\label{fig:crossover-long}     
   (a) Local single-particle Green function, (b) $\boldsymbol{q}=0$ current correlator,
   (c) identical charge and spin susceptibilities,
   and (d) slave-spin Green function of the FKM. Here, $L=8$. For clarity, (a),(b),(d)
   only show a subset of $\tau$ points.
 }
\end{figure}

The existence of a crossover between two distinct metallic regimes with O(4) symmetry
at $T>T_Q$ was established in Ref.~\RefI in terms of the results in
Figs.~\ref{fig:akw-z2} and~\ref{fig:crossover-long}. The single-particle
spectral function 
\begin{align}\label{eqn:akw}
  A(\bk,\omega)&=-\frac{1}{\pi}\mathrm{Im}\, G(\bk,\omega)
\end{align}
was extracted from the spin-averaged Green function
\begin{equation}\label{eq:greenfunction}
G(\bk,\tau) = \oh \sum_\sigma \la \oc^\dag_{\bk\sigma}(\tau)  \oc^\nag_{\bk\sigma}(0)\ra
\end{equation}
using a maximum entropy method \cite{Beach04a} to invert 
\begin{equation}
  G(\bk,\tau) = \int_{-\infty}^\infty \rmd \omega
  \frac{e^{-\omega\tau}}{1+e^{-\beta \omega}} A(\bk,\omega)\,.
\end{equation}
Figure~\ref{fig:akw-z2} reveals gapless excitations for
$U\lesssim 6$ (FL) but a gap at the Fermi level for $U\gtrsim 6$ (OM). The
absence (presence) of a single-particle gap at weak (strong) coupling can
also be inferred directly from the local imaginary-time Green function 
\begin{equation}\label{eq:localG}
G_{\text{loc}}(\tau)
= \frac{1}{L^{2}} \sum_{\bk} G(\bk,\tau)\,,
\end{equation}
which is related to the density of states
$N(\omega)$ by an identity akin to Eq.~(\ref{eq:greenfunction}). 
The Green function is clearly finite at long times $\tau$ for small $U$ but
exhibits an exponential decay determined by the single-particle gap at larger $U$.

At the same time, the  persistence of metallic behavior at large $U$ is
apparent from the current correlator
\begin{equation}\label{eq:currentcorrelator}
{\Gamma}_{xx}(\boldsymbol{q},\tau) = \frac{1}{L^2} \sum_{\bm{r}\bm{r}'}
e^{-\rmi(\bm{r}-\bm{r}')\cdot\bq}
\la\hat{\jmath}_x(\bm{r},\tau) \hat{\jmath}_x(\bm{r}',0)\ra 
\end{equation}
defined in terms of the current operator 
\begin{equation}\label{eq:currentoperator}
\hat{\jmath}_x(\bm{r}) = \rmi \sum_\sigma\left(
 \oc^\dag_{\bm{r}+\hat{e}_x,\sigma}\oc^\nag_{\bm{r}\vphantom{\hat{e}_x},\sigma}
- \oc^\dag_{\bm{r},\sigma\vphantom{\hat{e}_x}} \oc^\nag_{\bm{r}+\hat{e}_x,\sigma}
 \right)\,.
\end{equation}
Although $\Gamma_{xx}(\tau)={\Gamma}_{xx}(\boldsymbol{q}=0,\tau)$
in Fig.~\ref{fig:crossover-long}(b) also exhibits a significant suppression
with increasing $U$, it saturates at large $\tau$. This implies a Drude response
($\delta$ function at $T=0$, peak of finite width at $T>0$)
and hence a nonzero conductivity in the thermodynamic limit. A detailed
analysis of the (optical) conductivity and the resistivity will be given below.

A characteristic property of the OM regime is that despite the
single-particle gap, long-wavelength spin fluctuations remain gapless
(gapless charge fluctuations are naturally implied for a
metallic phase). This is visible from the uniform susceptibilities
\begin{equation}\label{eq:localsusc}
\chi_\alpha={\beta}\left(\las \hat{O}_\alpha^2\ras - \las\hat{O}^{}_\alpha\ras^2\right)
\end{equation}
for spin [$\hat{O}_\text{s}= \sum_i \hat{S}^z_i$, $\hat{S}^z_i=(\on_{i\UP}-\on_{i\DO})/2$]
and charge
($\hat{O}_\text{c} = \sum_i \on_i$), which are identical for the FKM due to
its O(4) symmetry. According to
Fig.~\ref{fig:crossover-long}(c), $\chi_\text{s}$ and $\chi_\text{c}$ are slightly enhanced in
the strong-coupling regime for $T>T_Q$, with a precursor effect
of the Ising transition visible at the lowest temperatures for $U=12$.
The curvature has a different sign in the FL and OM regime. Gapless spin and
charge excitations have also been demonstrated in terms of the dynamic
structure factor \RefI.

The evolution of the slave-spin Green function
\begin{equation}\label{eq:Gs}
G^s_\text{loc}(\tau)= \las \os^z_{i}(\tau)  \os^z_{i}(0)\ras
\end{equation}
with $U$ shown in Fig.~\ref{fig:crossover-long}(d) is very
similar to that of the electronic Green function~(\ref{eq:greenfunction}) in
Fig.~\ref{fig:crossover-long}(a). It reveals a crossover from frozen,
quasi-long-range correlations in imaginary time at small $U$
to short-ranged correlations at large $U$. The latter can be qualitatively 
captured in terms of a temporal ferromagnetic coupling
$K=J/T=\text{arctanh}(e^{-\Delta\tau U/4})$. The correlation length of the
corresponding 1D ferromagnetic Ising model is given by $(\ln\coth K)^{-1}\sim
1/U$. 

\subsection{Hubbard physics from the FKM}
\label{sec:results:hubbard}

\begin{figure}[t] 
\includegraphics[width=0.475\textwidth]{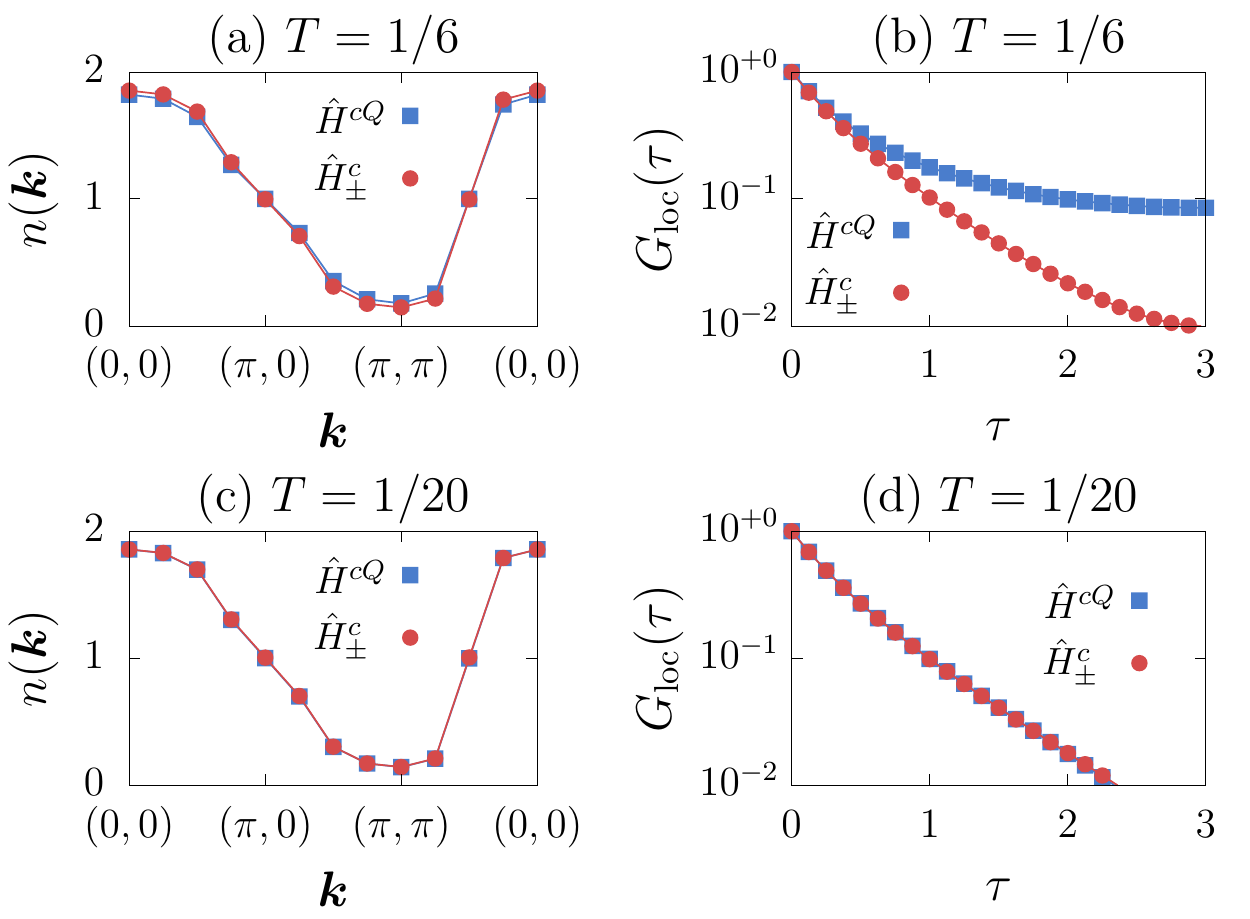}
 \caption{\label{fig:comparison-hubbard}  
 Comparison of (a), (c) the momentum distribution function and (b), (d) the local
 single-particle Green function for the FKM ($\hat{H}^{cQ}$) and the Hubbard
 model ($\hat{H}^{c}_\pm$) at (a), (b) high and (c), (d) low temperatures. Here, $U=6$, $L=8$.  
 }
\end{figure}

\begin{figure}[t]
  \includegraphics[width=0.475\textwidth]{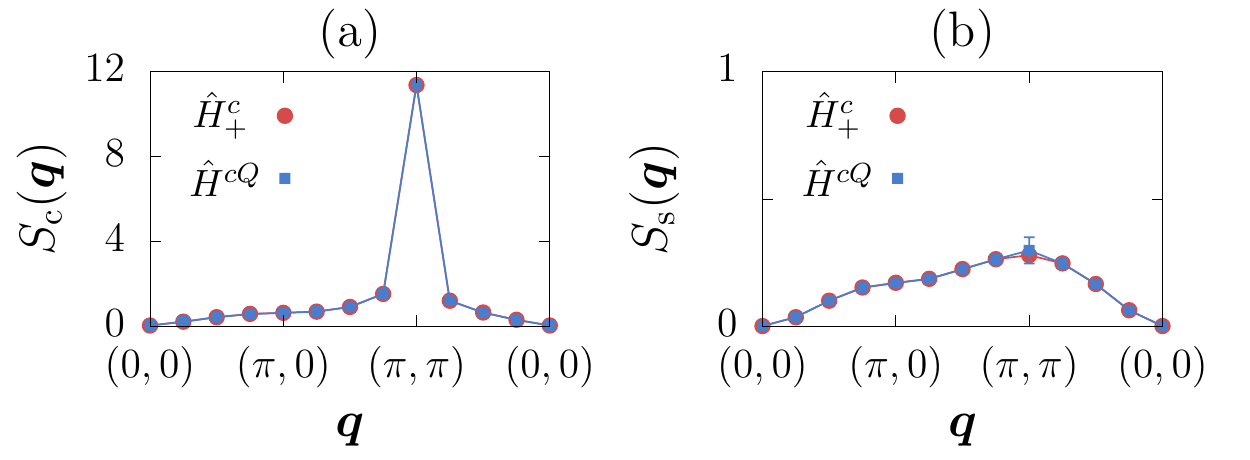}
  \caption{\label{fig:DenSpinU6} (a) Charge and (b) spin structure factor
    for the FKM ($\hat{H}^{cQ}$) with a magnetic field $h_Q=0.05$ and the attractive Hubbard
    model ($\hat{H}^{c}_-$). Here, $T=0.05$, $U=6$, and $L=8$.
  }
\end{figure}

According to Sec.~\ref{sec:slavespin} and Fig.~\ref{fig:grandphasediagram}, the FKM~(\ref{eq:HcQ})
should exhibit quantitative agreement with the Hubbard model~(\ref{eq:hubbard})
in several parameter regions. To obtain explicit numerical evidence, we have to keep in mind that the
spontaneous breaking of the relevant \ZII symmetry only occurs in the
thermodynamic limit.  Our simulations in the representation~(\ref{eq:Hftau})
amount to a perfect sampling of the $\oQ_i$, leading to an average
over pairs of $\oQ_i$ configurations related by a global spin flip $\oQ_i\to-\oQ_i$.
Based on the discussion in Sec.~\ref{sec:slavespin} and Eq.~(\ref{eq:HcQ}),
this implies an average over the attractive and 
repulsive sectors of the Hubbard model that produces the O(4) symmetry.
There are two routes to verify agreement with the Hubbard model in
finite-size simulations: (i) considering observables that are
invariant under the transformation~(\ref{eq:phtrafo}) and hence identical for
$U>0$ and $U<0$, (ii) breaking the O(4) symmetry explicitly.

Regarding the first possibility, we show in Fig.~\ref{fig:comparison-hubbard}
the momentum distribution function 
\begin{equation}\label{eq:nofk}
n(\bk)=\sum_\sigma \la \oc^\dag_{\bk\sigma}\oc^\nag_{\bk\sigma}\ra
\end{equation}
and the local Green function defined in Eq.~(\ref{eq:localG}), which depend only on $|U|$. We take
$U=6$, close to the maximum of $T_Q\approx0.1$ in
Fig.~\ref{fig:phasediagram}. While quantitative
differences between the FKM and the Hubbard model are
visible in both $n(\bk)$ and $G_\text{loc}(\tau)$ at $T=1/6> T_Q$ [Figs.~\ref{fig:comparison-hubbard}(a) and (b)],
results are essentially identical at $T=0.05< T_Q$
[Figs.~\ref{fig:comparison-hubbard}(c) and (d)].

By adding a small magnetic field $h_Q$ for the $\oQ_i$ or, equivalently, a small
Hubbard interaction of strength $U'=h_Q$ [cf. Eq.~(\ref{eq:hubbard})], we can explicitly break the O(4) symmetry. This
leads to quantitative agreement on finite systems, as illustrated in Fig.~\ref{fig:DenSpinU6} for the charge and
spin structure factors
\begin{align}\label{eq:structurefactors}\nonumber
S_\text{c}(\bq) &= \frac{1}{L^2}\sum_{ij} e^{-\rmi(\bm{r}_i-\bm{r}_j)\cdot\bq}\la (\on_i-1)(\on_j-1)\ra \,,\\
S_\text{s}(\bq) & = \frac{1}{L^2}\sum_{ij} e^{-\rmi(\bm{r}_i-\bm{r}_j)\cdot\bq}\la \hat{S}^z_i\hat{S}^z_j\ra\,,
\end{align}
A field $h_Q=0.05$
is sufficient to select the ferromagnetic phase with $\oQ_i=1$ and thereby
generate results that are identical to those for the attractive Hubbard model.

\begin{figure}[t] 
\includegraphics[width=0.475\textwidth]{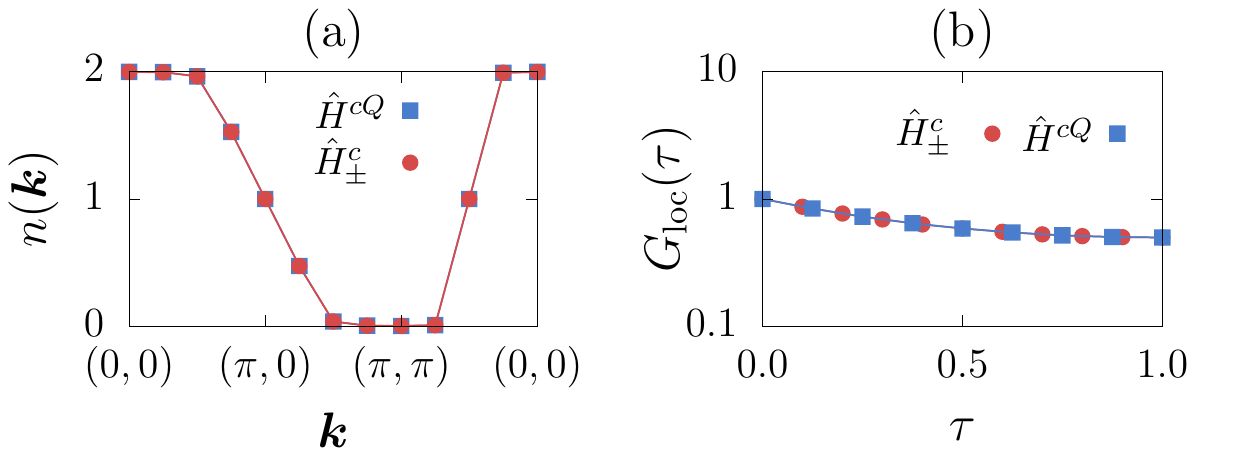}
 \caption{\label{fig:comparison-hubbard-0.1}  
   Comparison of (a) the momentum distribution function and (b) the local
   Green function for the FKM ($\hat{H}^{cQ}$) and the Hubbard
   model ($\hat{H}^{c}_\pm$). Here, $U=0.1$, $T=0.5$, and $L=8$. 
 }
\end{figure}

Finally, Fig.~\ref{fig:comparison-hubbard-0.1} demonstrates the asymptotic
irrelevance of the constraints near $U=0$ {\em at any temperature} in terms
of excellent agreement for the same observables as in
Fig.~\ref{fig:comparison-hubbard} at a temperature $T=0.5\gg T_Q$.

\subsection{Metallic regimes at $T>T_Q$}\label{sec:results:FL-OM}

At $T>T_Q$, the constraints $\oQ_i$ are disordered, so that the FKM yields
physics beyond the Hubbard model. Here, we first consider the crossover from the gapless FL
to the gapped OM with increasing $U$, before turning to the
temperature dependence.

\subsubsection{Crossover from weak to strong coupling}

\begin{figure}[t] 
\includegraphics[width=0.475\textwidth]{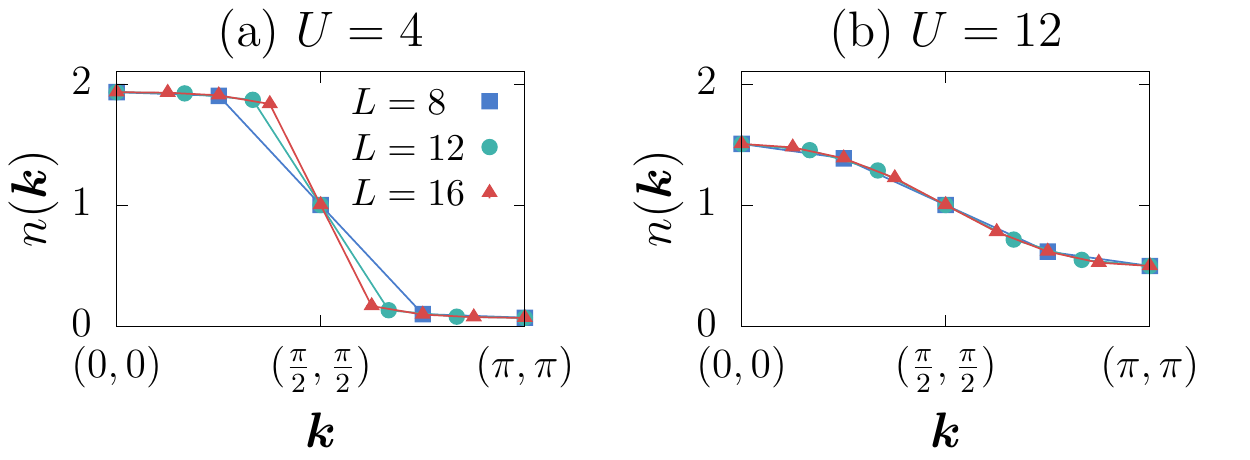}
 \caption{\label{fig:nk-FSS}  
   Momentum distribution function of the FKM at $T=1/6$ in
   (a) the FL and (b) the OM regime.
 }
\end{figure}

A crossover from the FL regime with well-defined quasiparticles to the OM
regime without quasiparticles is suggested by Fig.~\ref{fig:akw-z2}.
To substantiate this observation, we show in Fig.~\ref{fig:nk-FSS} the
momentum distribution function [Eq.~(\ref{eq:nofk})] for different system sizes.
While a true jump can only exist at $T=0$, the results for the FL regime at
$U=4$ in Fig.~\ref{fig:nk-FSS}(a) are consistent with the existence of quasiparticles.
In contrast, in the OM regime at $U=12$ [Fig.~\ref{fig:nk-FSS}(b)], the
continuous dependence of $n(\bk)$ on $\bk$ across the Fermi surface with 
essentially no finite-size effects supports the absence of quasiparticles. 

Previously \RefI, we calculated the dc conductivity using the estimator \cite{PhysRevB.54.R3756}
\begin{equation}\label{eq:sigmafromJJ}
  \sigma_\text{dc}^{(1)}
  \approx
  \frac{\beta^2}{\pi}
  {\Gamma}_{xx}(\beta/2)
\end{equation}
(results in Ref.~\RefI include an additional factor $2$ from the trace ${\Gamma}_{xx} + \Gamma_{yy}$). An improved estimator that
takes into account the curvature of ${\Gamma}_{xx}$ at $\tau=\beta/2$ is \cite{lederer2017superconductivity}
\begin{equation}\label{eq:sigmafromJJ2}
  \sigma_\text{dc}^{(2)}
  \approx
  \frac{2\pi [{\Gamma}_{xx}(\beta/2)]^2}
  {\partial^2_\tau{\Gamma}_{xx}(\beta/2)} 
\end{equation}
and involves a quadratic fit near $\tau=\beta/2$
\cite{lederer2017superconductivity} (here, statistical errors were propagated
but we did not quantify the impact of the fit range). 

\begin{figure}[t] 
\includegraphics[width=0.475\textwidth]{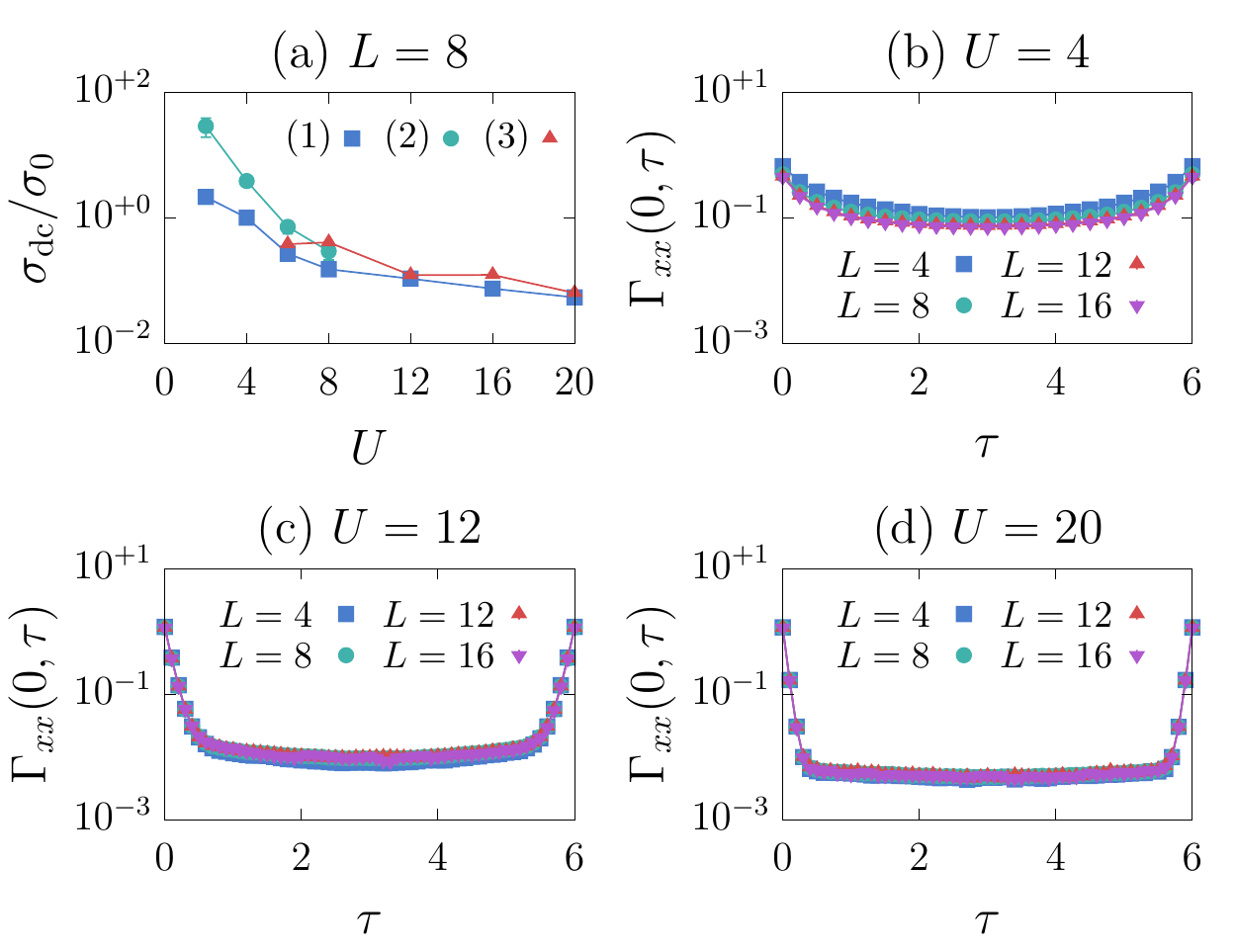}
\caption{\label{fig:crossover_dc}  
  (a) Conductivity of the FKM from Eqs.~(\ref{eq:sigmafromJJ})\,--\,(\ref{eq:sigmafromMEM}).
  (b)\,--\,(d) Current correlator for different $L$. Here, $T=1/6$.
 }
\end{figure}

\begin{figure}[t] 
\includegraphics[width=0.475\textwidth]{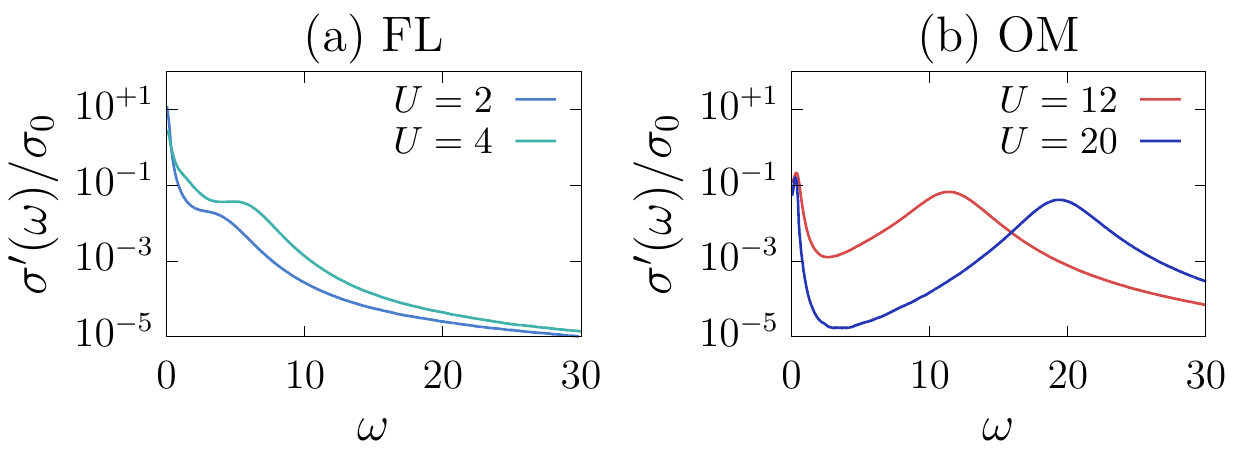}
 \caption{\label{fig:optcond_Z2_U}     
   Optical conductivity of the FKM for $T=1/6$ and $L=8$. Here and in subsequent figures showing $\sigma'(\om)$, the energy
   resolution is $0.02$ and we applied a cubic Savitzky-Golay filter to
   obtain smooth(er) curves.}
\end{figure}

We also extracted $\sigma_\text{dc}$ from the real part of the
optical conductivity $\sigma'(\omega)$ via 
\begin{equation}\label{eq:sigmafromMEM}
\sigma_\text{dc}^{(3)}=\lim_{\omega\to 0} \sigma'(\omega)\,.
\end{equation}
This required a maximum-entropy inversion \cite{Beach04a} of the Laplace transform 
\begin{equation}
  {\Gamma}_{xx}(\tau) = \int 
  \frac{\rmd \omega}{\pi}
  \frac{e^{-\tau\omega}}{1 - e^{-\beta\omega}}
  \omega \,\sigma'(\omega) \,.
\end{equation}
The f-sum rule for the optical conductivity reads
\cite{PhysRevB.16.2437,PhysRevX.4.021007}
\begin{equation}\label{eq:fsumrule}
2\int_{0^+}^\infty \rmd\omega\, \sigma'(\omega) = -\pi e^x_{\text{kin}} - \rho_\text{s}
\end{equation}
with the kinetic energy per bond ($t=1$)
\begin{equation}\label{eq:ekinperbond}
  e^x_{\text{kin}}   
  =
  \frac{1}{2L^2}\sum_{\boldsymbol{r}\sigma} 
  \las \oc^\dag_{\boldsymbol{r}\vphantom{e_x}\sigma}\oc^\nag_{\boldsymbol{r}+\hat{e}_x,\sigma} + \text{H.c.}\ras
\end{equation}
and the superfluid density $\rho_\text{s}$ defined in
Eq.~(\ref{eq:superfluid1}). We find that Eq.~(\ref{eq:fsumrule}) holds 
numerically as long as $\sigma'(\omega)$ does not have a very sharp
low-energy feature that is nontrivial to reliably extract with the
maximum-entropy method. Results for $\sigma'(\om)$ are shown despite
non-negligible deviations (more
than five percent) if only the qualitative structure is of interest. This
applies to $U=2,4$ in Fig.~\ref{fig:optcond_Z2_U}(a) and to $T=1/6$ in
Fig.~\ref{fig:optcond_beta}(a). On the other hand, such data  
are excluded from the quantitative analysis of the dc conductivity in
Figs.~\ref{fig:crossover_dc}(b) and~\ref{fig:resistivity-fkm}.

The accuracy of the estimators~(\ref{eq:sigmafromJJ}) and~(\ref{eq:sigmafromJJ2})
depends on the spectral shape of $\sigma'(\omega)$ and the temperature, see
Refs.~\cite{PhysRevB.54.R3756,lederer2017superconductivity,huang2018strange}.
For example, $\sigma^{(1)}_\text{dc}$ gives reliable results if
$\sigma'(\omega)$ has no low-energy contribution sharper than $\Delta\omega\approx 8T$
\cite{huang2018strange}. The estimator~(\ref{eq:sigmafromJJ2}) varies between
$\sigma_\text{dc}$ for a single low-energy peak of width $\Delta\omega\gg T$ and
$\sigma_\text{dc}/2$ for $\Delta\omega\ll T$ \cite{huang2018strange}, but larger
deviations are possible if $\sigma'(\omega)$ has a different form. Similarly,
$\sigma_\text{dc}$ values extracted from the maximum-entropy results for $\sigma'(\omega)$
are subject to uncertainties that are challenging to quantify.
At the same time, for the doped Hubbard model at high temperatures,
qualitative or even quantitative agreement of different estimators was
observed \cite{huang2018strange}.

Here, the simplest but least accurate estimator~(\ref{eq:sigmafromJJ})
can be calculated for almost arbitrary parameters. In contrast, the use of
Eq.~(\ref{eq:sigmafromJJ2}) becomes challenging if  $\Gamma_{xx}$ is flat
around $\beta/2$, as is the case for large $U$ [see Fig.~\ref{fig:crossover-long}(b)].
Overall, while we cannot expect quantitative or even qualitative
agreement, a comparison of all three estimators should reveal systematic trends.

Figure~\ref{fig:crossover_dc}(a) shows $\sigma_\text{dc}$ as a
function of $U$ at $T=1/6$. The results reaffirm the findings of Ref.~\RefI: the conductivity
drops sharply in the gapless weak-coupling regime, but much more slowly in
the strong-coupling OM regime. At small $U$,  the fact that $\sigma^{(1)}_\text{dc}$
underestimates the conductivity can be attributed to the significant
variation of $\sigma'(\omega)$ over $\omega\in[0,8T]$. In the OM regime, all
three estimators yield quantitatively similar results. In total, we therefore
see an even more pronounced difference between weak and strong coupling than in Ref.~\RefI.

To rule out that the nonzero conductivity in the OM regime is
merely a finite-size effect, Figs.~\ref{fig:crossover_dc}(b)\,--\,(d) show
the current correlation function for different system sizes $L$ in the FL and
the OM regimes. Even in the FL regime [Fig.~\ref{fig:crossover_dc}(b)],
finite-size effects are minimal for the temperatures considered. In the OM
regime, results for different $L$ are essentially identical and hence support
a nonzero conductivity in the thermodynamic limit. This is of particular
interest in connection with the 2D spinless FKM, for which metallic behavior
has recently been argued to be absent at any $U>0$ for $L\to\infty$
\cite{PhysRevLett.117.146601}, see Sec.~\ref{sec:discussion}.

Figure~\ref{fig:optcond_Z2_U} reveals the impact of the FL-OM crossover on
the optical conductivity itself. At small $U$, where the single-particle spectrum is
gapless [Figs.~\ref{fig:akw-z2}(a)--(b)], $\sigma'(\omega)$ in
Fig.~\ref{fig:optcond_Z2_U}(a) exhibits a dominant zero-frequency, Drude type
contribution as well as a second peak centered at $\omega\approx U$. A reduced
Drude response remains in the OM regime, Fig.~\ref{fig:optcond_Z2_U}(b),
reflecting its metallic nature. Both the position ($\om\sim U$) and height of 
the `Hubbard peak' remain largely unchanged. A significant
shift of weight from zero to finite frequencies when comparing the weak
and strong-coupling regimes is typical of strongly correlated systems \cite{PhysRevLett.75.105}. The high-frequency peak is
expected from the gap in the single-particle spectrum in Fig.~\ref{fig:akw-z2}. On the other hand, the
low-frequency peak is entirely beyond a local theory without vertex
corrections where a gap in $A(\bk,\om)$ also implies a gap in
$\sigma'(\omega)$ \cite{PhysRevB.47.3553}. 

\begin{figure}[t] 
\includegraphics[width=0.475\textwidth]{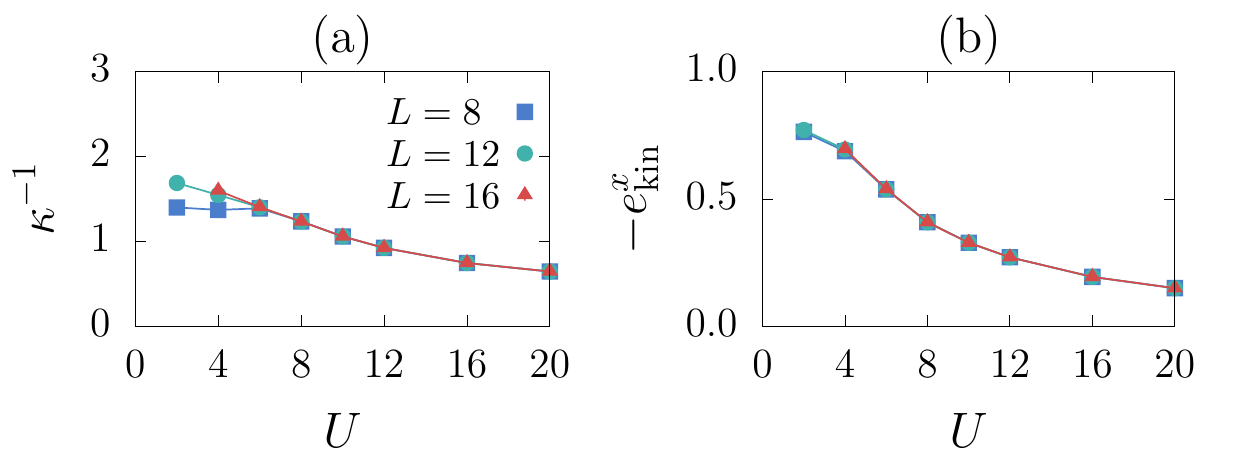}
 \caption{\label{fig:compress-vs-U} 
   (a) Inverse compressibility and (b) resistivity from
   Eq.~(\ref{eq:sigmafromJJ}) of the FKM for different system sizes. Here, $T=1/6$.}
\end{figure}

An absence of critical behavior, in accordance with a crossover rather than
a phase transition separating the FL and OM regimes, can be seen in
Fig.~\ref{fig:compress-vs-U}. The inverse compressibility and the
kinetic energy both vary smoothly with $U$ and exhibit only a very weak
dependence on the system size $L$.

\subsubsection{Absence of superconductivity}

While we interpret the gapped metallic regime at large $U$ in the
framework of OMs, an apparent alternative explanation is superconductivity. The
latter can be ruled out  by calculating the superfluid density
\cite{PhysRevB.47.7995}
\begin{equation}\label{eq:superfluid1}
 {\rho_\text{s}} = \pi \left[- e^x_{\text{kin}} - \lim_{q_y\to0}\Gamma_{xx}(q_y)\right]\,.
\end{equation}
Here, $\Gamma_{xx}(q_y)\equiv\Gamma_{xx}(q_x=0,q_y,\omega=0)$ corresponds to the
integral of the current correlator defined in Eq.~(\ref{eq:currentcorrelator}) over $\tau$.
In a normal state, it is identical to the kinetic energy [Eq.~(\ref{eq:ekinperbond})] 
so that $\rho_\text{s}=0$. Superconductivity implies a nonzero
difference $\rho_\text{s}>0$ that has to be accounted for in the f-sum
rule~(\ref{eq:fsumrule}) \cite{PhysRevX.4.021007}.

\begin{figure}[t] 
\includegraphics[width=0.475\textwidth]{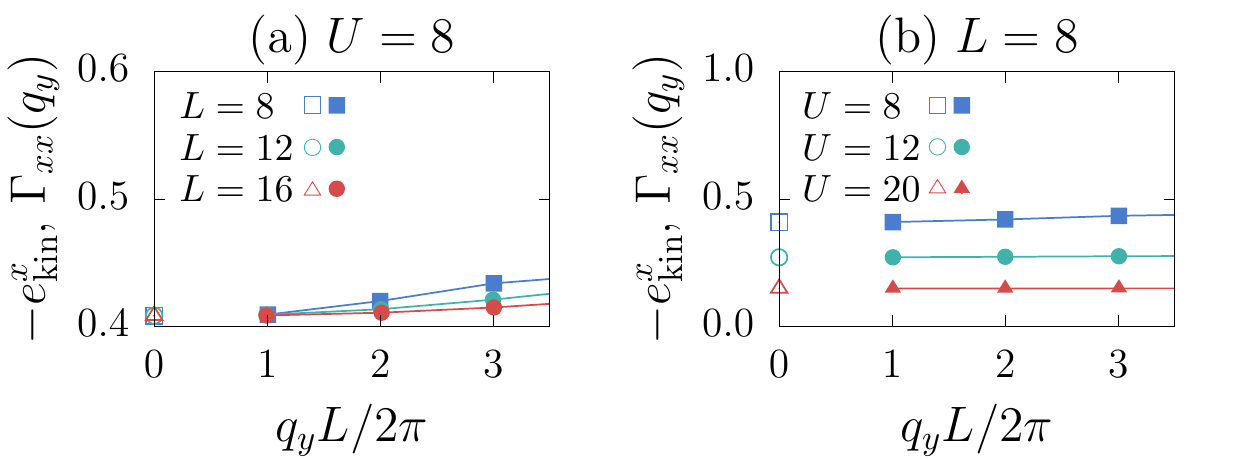}
 \caption{\label{fig:superfluid} 
Transverse current response (filled symbols) and negative of the kinetic energy
   (open symbols) of the FKM in the OM regime at $T=1/6$.
}
\end{figure}

Figure~\ref{fig:superfluid}(a) shows that $\Gamma_{xx}(q_y)$ approaches
$-e^x_{\text{kin}}$ in the OM regime ($U=8$, $T=1/6$), indicating the absence
of superconductivity. The same conclusion holds for larger values of $U$, as
demonstrated in Fig.~\ref{fig:superfluid}(b). Different behavior will be
discussed below for the attractive Hubbard model.

\subsubsection{Temperature dependence}

\begin{figure}[t] 
   \includegraphics[width=0.45\textwidth]{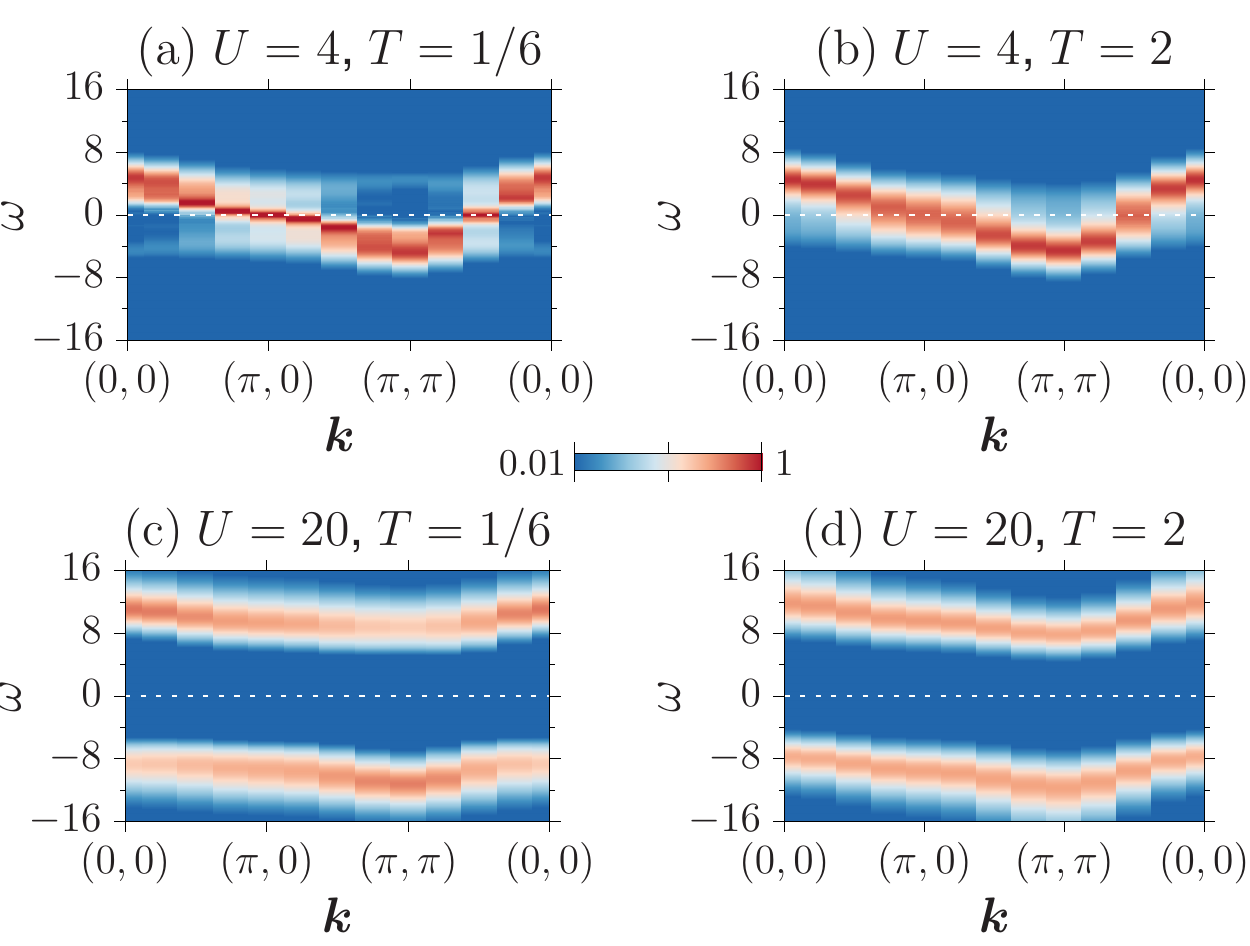}   
  \caption{\label{fig:akw-temp-U12} Single-particle spectral function
    of the FKM in (a),(b) the FL and (c),(d) the OM regime. Here, $L=8$.
  }
\end{figure}
\begin{figure}[t] 
   \includegraphics[width=0.475\textwidth]{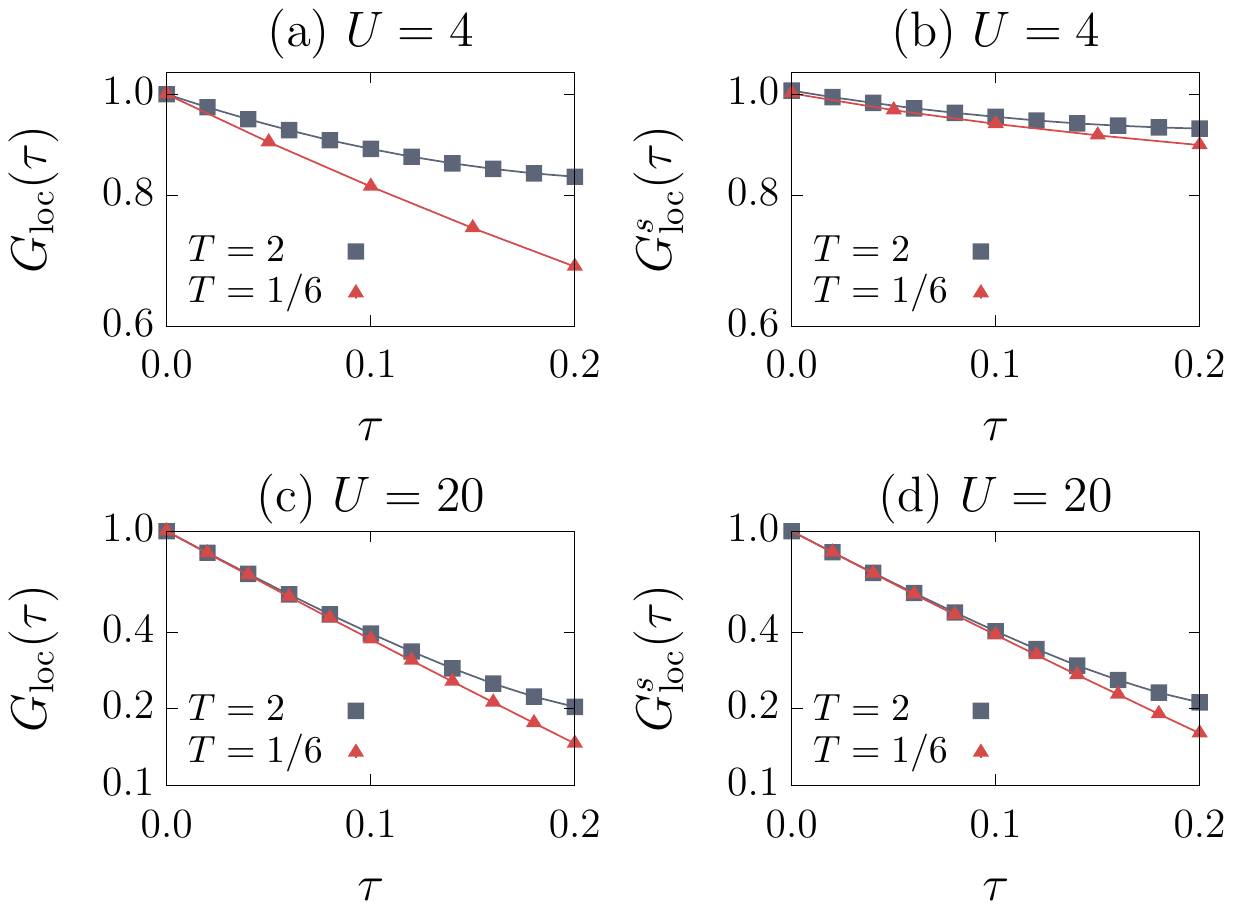}     
  \caption{\label{fig:gf-temp-U12} (a),(c) Local electron Green function and (b),(d)
    slave-spin Green function of the FKM for $L=8$.}
\end{figure}

\begin{figure}[t] 
\includegraphics[width=0.45\textwidth]{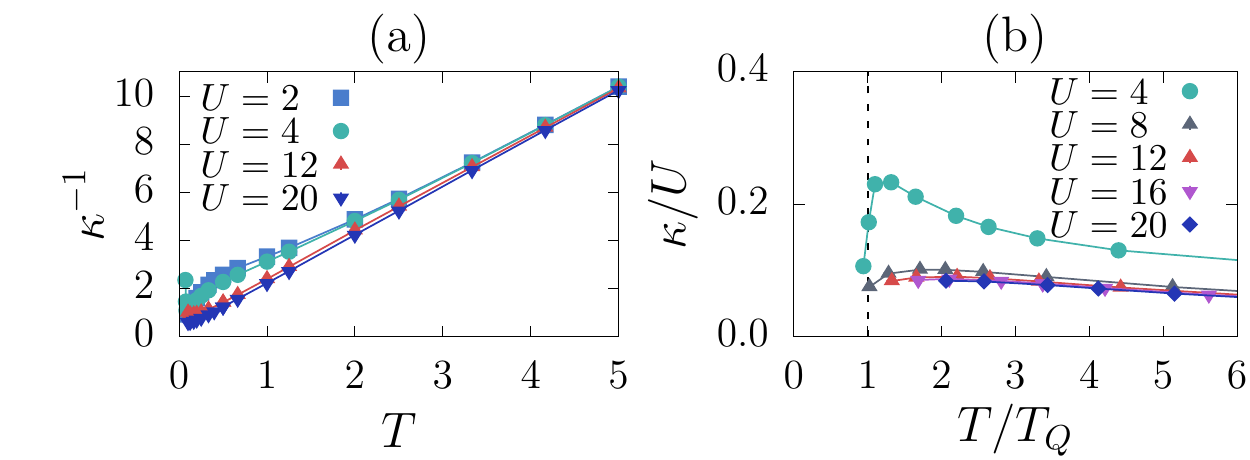}
 \caption{\label{fig:compress-fkm}  (a) Inverse compressibility of the FKM
   as a function of temperature. 
   (b) Rescaled compressibility as a function of $T/T_Q$ with $T_Q$ from Fig.~\ref{fig:phasediagram}.
   Here, $L=8$.}
\end{figure}

The temperature dependence of quantities such as the resistivity and the
optical conductivity is of particular interest in connection with
experiments. For example, it distinguishes FLs from more exotic bad or strange
metals \cite{hussey2004universality}. For the FKM~(\ref{eq:HcQ}),
Fig.~\ref{fig:phasediagram}  implies that regardless of methodological
restrictions, the Ising transition at $T_Q$ does not permit us to track the
temperature dependence of the OM all the way to $T=0$.

Figure~\ref{fig:akw-temp-U12} compares the spectral function $A(\bk,\omega)$
at low ($T=1/6$) and high temperatures ($T=2$) in the FL and OM regimes.
Whereas for $U=4$ (FL) the gapless excitations become significantly
broadened around the Fermi level, the spectrum remains virtually unchanged or
becomes even slightly sharper at high temperatures for $U=20$ (OM).

The corresponding local Green functions of the $c$ fermions and the slave
spins are shown in Fig.~\ref{fig:gf-temp-U12}. In the FL regime, the impact of
thermal fluctuations is significantly larger for the electrons
[Fig.~\ref{fig:gf-temp-U12}(a)] than for the slave spins
[Fig.~\ref{fig:gf-temp-U12}(b)] in terms of imaginary-time correlations at
large $\tau$ that determine the gapless excitations in Fig.~\ref{fig:akw-temp-U12}(a).
This suggests that in the FL, the quasi-ordered slave spins do not fully determine the electronic excitations. In contrast,
in the OM regime, both sectors exhibit very similar imaginary-time correlations
and temperature effects. The robustness of the gapped, high-energy bands in
$A(\bk,\omega)$ is reflected in an almost complete absence of thermal effects
at small $\tau$ in Figs.~\ref{fig:gf-temp-U12}(c) and (d). This can be attributed to the fact that the transverse field
$h=U/4$  is much larger than the thermal energy $T$. 

The charge susceptibility defined in Eq.~(\ref{eq:localsusc}) and
shown in Fig.~\ref{fig:crossover-long}(c) is identical to the
compressibility
\begin{equation}
\kappa 
 = \frac{1}{\las n\ras^2}\frac{\partial \las n\ras}{\partial \mu}
 \,,
\end{equation}
as can be readily established by differentiating the total particle number $N
= L^2 \las n\ras =\tr (\hat{N} e^{-\beta (\hat{H}-\mu\hat{N})})/Z$ and
setting $\las n\ras=1$. Independent of $U$, the inverse compressibility in
Fig.~\ref{fig:compress-fkm}(a) increases with temperature at high temperatures.
However, as visible in Fig.~\ref{fig:crossover-long}(c), the FL and OM
regimes are characterized by different signs of the curvature of $\kappa^{-1}$ at low temperatures. For $T>T_Q$, the
compressibility  is larger ($1/\kappa$ is smaller) in the OM than in the
FL regime. The linear dependence at high temperatures can be understood in
terms of the Curie behavior of quasi-free
fermions.

Figure~\ref{fig:compress-fkm}(a) shows $\kappa/U$ as a function of
$T/T_Q$. As in a DMFT study of the same model
\cite{tran2018fractionalization}, we find a significant dependence on $U$ in
the FL regime but an approximate collapse of results for different $U$ in the
OM regime. For $U=4$, where results are shown also for $T\lesssim T_Q$,
Fig.~\ref{fig:compress-fkm}(b) reveals a pronounced drop at the critical
temperature (\ie, at $T/T_Q=1$). Hence, the Ising phase transition of the
$\oQ_i$, corresponding to a spontaneous symmetry reduction
O(4)$\mapsto$SO(4), has a clear electronic signature, even though
long-range order in the fermionic sector exists only at $T=0$. This is in
strong contrast to the much smoother evolution of $\kappa^{-1}$ as a function of
$U$ in Fig.~\ref{fig:compress-vs-U}(a), consistent with an FL-OM crossover
rather than a phase transition.

\begin{figure}[t] 
\includegraphics[width=0.45\textwidth]{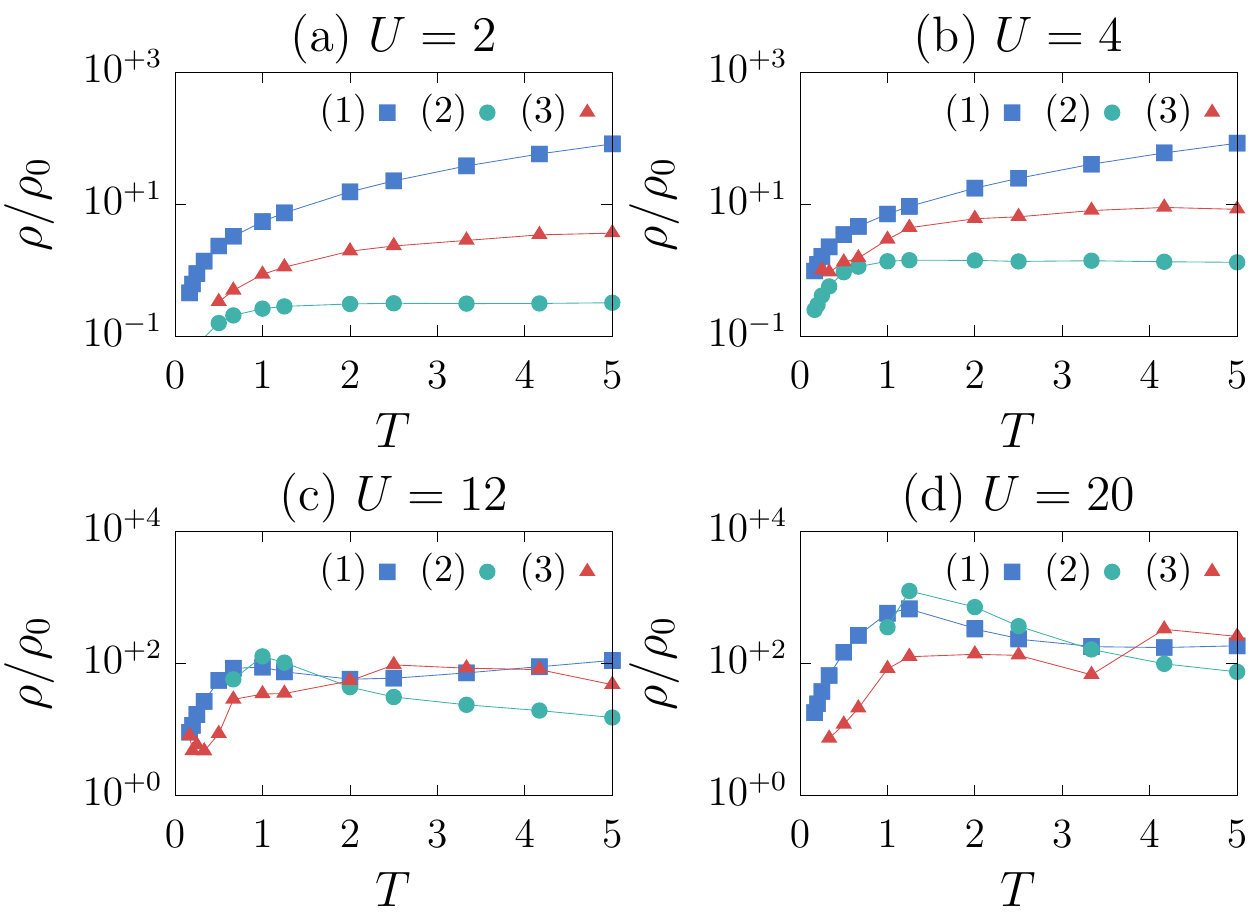}
\caption{\label{fig:resistivity-fkm} Resistivity of the FKM from the different
  estimators $\rho^{(i)}=1/\sigma_\text{dc}^{(i)}$ in (a), (b) the FL regime 
  and (c), (d) the OM regime. Here, $L=8$. 
}
\end{figure}

Using the different conductivity estimators to calculate the resistivities
$\rho^{(i)}=1/\sigma^{(i)}_\text{dc}$ reveals a larger variation among
estimators than in Fig.~\ref{fig:crossover_dc}(b). Nevertheless, the results
in Fig.~\ref{fig:resistivity-fkm} exhibit interesting common trends. In 
the FL regime [Fig.~\ref{fig:resistivity-fkm}(a)--(b)], $\rho$ increases strongly with temperature at low
temperatures. For $T\gtrsim 1$, $\rho^{(1)}$ increases much more slowly 
whereas the other estimators suggest a saturating resistivity.
Two distinct temperature regimes are also visible in the OM regime
[Fig.~\ref{fig:resistivity-fkm}(c)--(d)]. Here, $\rho$ is overall
significantly larger than in the FL regime, in accordance with 
the conductivity in Fig.~\ref{fig:crossover_dc}(b).
The resistivity increases strongly with $T$ at low temperatures.
The estimators  $\rho^{(1)}$ and $\rho^{(2)}$ exhibit a maximum
around $T=1$. Whereas $\rho^{(1)}$ remains essentially unchanged in the range
$1\lesssim T \lesssim 5$, $\rho^{(2)}$ decreases with increasing temperature,
suggesting a potential crossover from metallic to insulating behavior.
The maximum at $T\approx 1$ is more pronounced for $U=20$ than for $U=12$.
Finally, the maximum-entropy estimator $\rho^{(3)}$ is also consistent with a
strong increase at low temperatures and saturation at high temperatures.
The additional crossover from linear to quadratic
behavior at low temperatures $T<0.5$ discussed in Ref.~\RefI cannot be addressed
by estimators $(2)$ and $(3)$. Whereas the present results reveal saturation
or even a decrease of the resistivity in the OM regime, a non-saturating
resistivity and inverse compressibility were observed for the doped Hubbard
model at comparable temperatures \cite{PhysRevB.94.235115,huang2018strange}.

\begin{figure}[t]
  \includegraphics[width=0.475\textwidth]{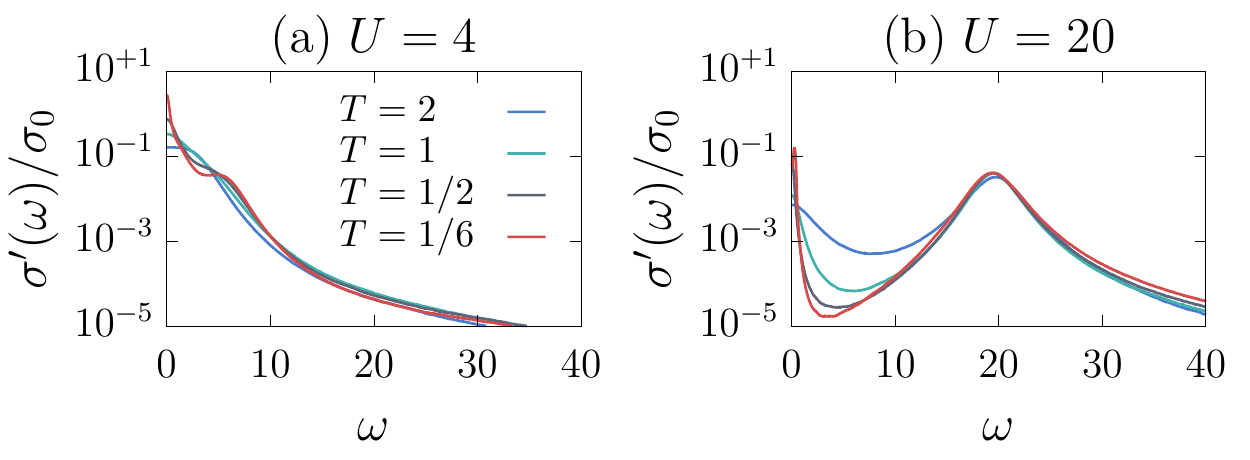} 
  \caption{\label{fig:optcond_beta} Optical conductivity of the FKM for
    (a) the FL regime and (b) the OM regime. Here, $L=8$.
  }
\end{figure}

The evolution of the optical conductivity with increasing $T$ can be seen in
Fig.~\ref{fig:optcond_beta}. In the FL regime at $U=4$
[Fig.~\ref{fig:optcond_beta}(a)], the two separate peaks visible at low
temperatures merge into a single peak at high temperatures, mainly due to the
broadening of the Drude peak with increasing $T$. In contrast, in the OM
regime at $U=20$ shown in Fig.~\ref{fig:optcond_beta}(b), despite the
filling-in of the spectral gap, two separate peaks remain clearly visible. 
The upper peak at $\om\sim U$  shows only a minor shift to smaller
frequencies with increasing $T$, in accordance with the slight reduction of
the gap in Fig.~\ref{fig:akw-temp-U12}. The ratio of spectral weights
contained in the lower and upper peaks changes from about 1:4 at $T=1/6$ to 1:8 at $T=2$.

\begin{figure}[t] 
\hspace*{0.5em}
\includegraphics[width=0.45\textwidth]{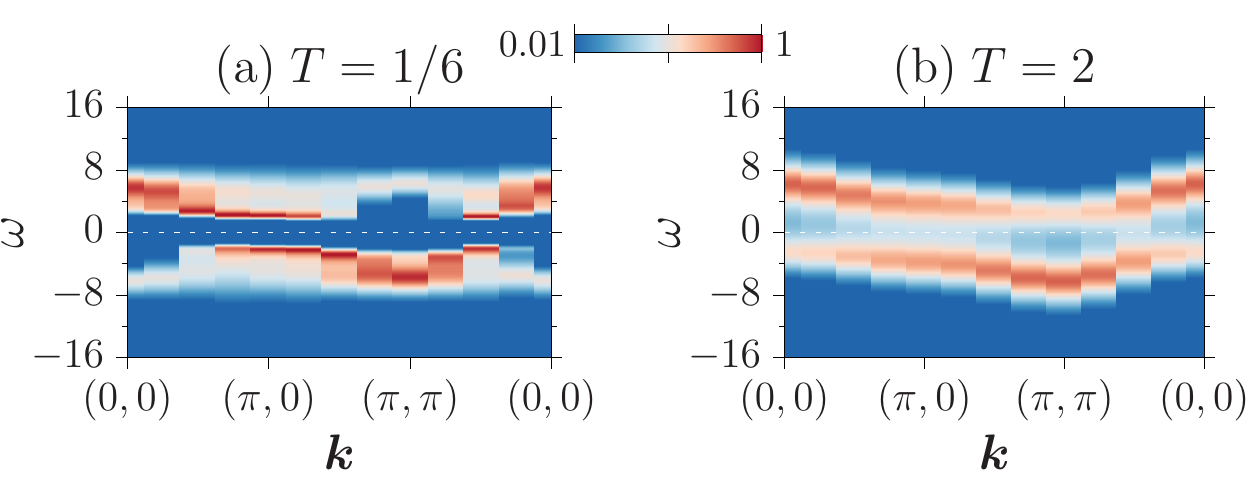}\\
\includegraphics[width=0.475\textwidth]{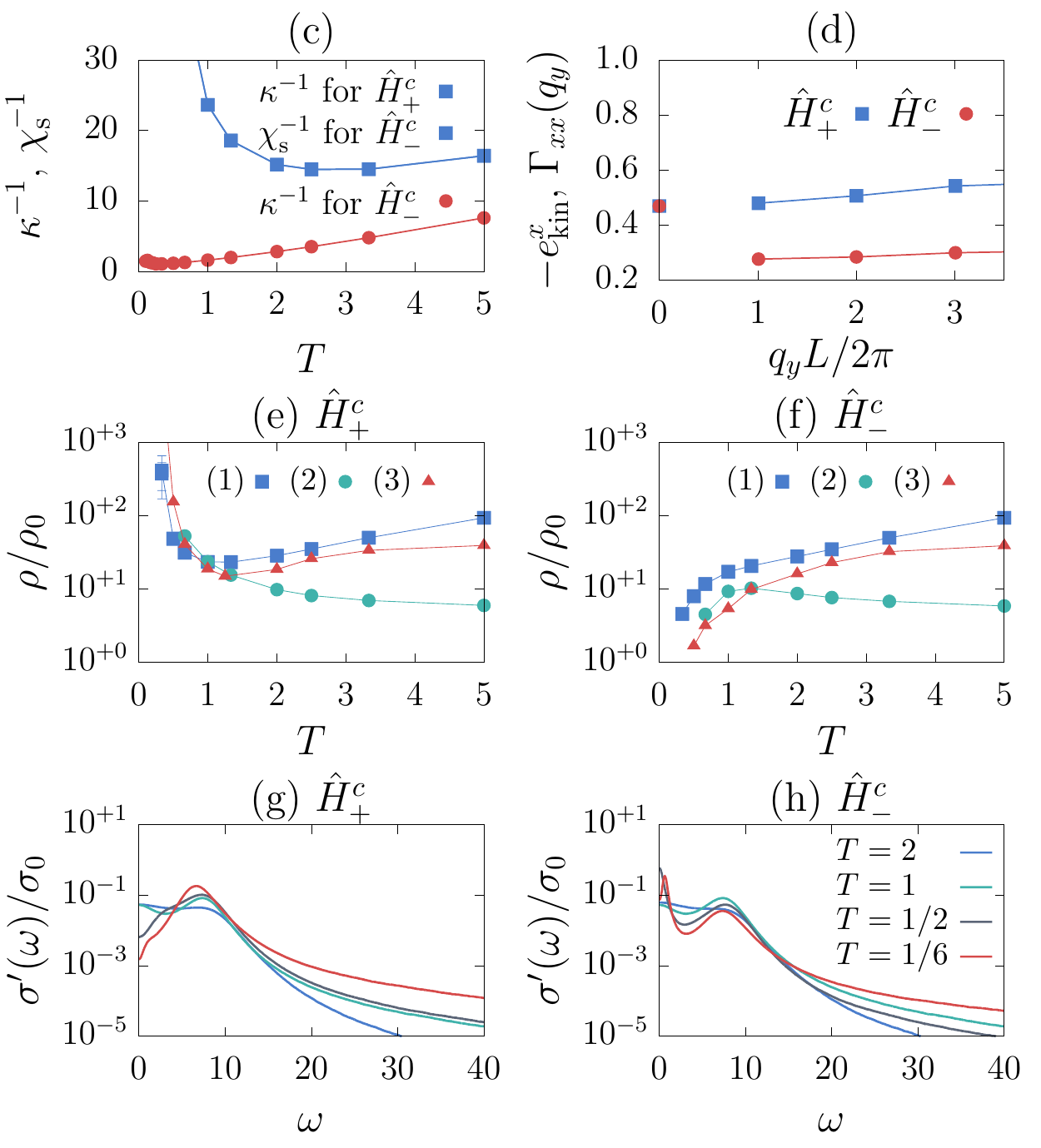}
 \caption{\label{fig:resistivity-hubbard} 
   QMC results for attractive ($\oH^c_-$) and repulsive ($\oH^c_+$) Hubbard
   models. (a)--(b) Single-particle spectral function at (a) low and
   (b) high temperature, independent of the sign of $U$. (c) Inverse
   compressibility or (upon $\oH^c_-\leftrightarrow\oH^c_+$) inverse spin susceptibility, (d) transverse current response and kinetic energy,
   (e)--(f) resistivity from the three different estimators, and (g)--(h)
   optical conductivity for different temperatures.
   Here, $U=8$ and $L=8$.
}
\end{figure}

\subsection{Differences to the half-filled Hubbard model}\label{sec:delineate-hubbard}

Given the relation of the FKM to the half-filled Hubbard model, we provide a
brief comparison. To this end, we consider both attractive ($\oH^c_{-}$) and 
repulsive ($\oH^c_{+}$) interactions. The ground state changes qualitatively
with the sign of the interaction but depends only quantitatively on the magnitude.
Here, we focus on $U=8$, for which the relevant temperature and energy scales are
readily accessible in QMC simulations.

At half-filling, the single-particle spectral function is independent of the
sign of $U$. Its temperature dependence has been
studied in detail before~\cite{PhysRevB.62.4336}. For $T$ above the magnetic scale
$J=4t^2/U=0.5$ [Fig.~\ref{fig:resistivity-hubbard}(b)], 
the spectrum exhibits  upper and lower Hubbard bands separated by a gap. In
contrast, at  $T<J$ [Fig.~\ref{fig:resistivity-hubbard}(a)], $A(k,\omega)$ has an
intricate four-band structure not captured by the Hubbard I
approximation. The additional features can be linked to the dressing of
electron and hole excitations with coherent spin waves
 that emerge at low temperatures \cite{PhysRevB.62.4336}. This can be
contrasted with the FKM [Fig.~\ref{fig:akw-temp-U12}(d)], where such
physics is absent at $T>T_Q$.

The inverse compressibility in Fig.~\ref{fig:resistivity-hubbard}(c) reveals
the suppression of charge fluctuations for repulsive interactions, as
opposed to the attractive case.  For the latter, the
behavior of $\kappa^{-1}$ is qualitatively similar to the OM regime,
cf. Fig.~\ref{fig:compress-fkm}(b). By the transformation~(\ref{eq:phtrafo}),
$\kappa^{-1}$ for the repulsive (attractive) Hubbard model is also identical to
the uniform spin susceptibility  of the attractive (repulsive) Hubbard model. Hence,
Fig.~\ref{fig:resistivity-hubbard}(c) reveals an exponential decrease of 
$\chi_\text{s}$ of the attractive model due to spin gap formation, whereas
spin excitations remain gapless for repulsive interactions. The spin gap
is perhaps the most pronounced difference between the attractive Hubbard
model at $T>0$ and the FKM in the OM regime [Fig.~\ref{fig:crossover-long}(c)].

In contrast to the FKM (Fig.~\ref{fig:superfluid}), the attractive 
Hubbard model exhibits signatures of superconductivity (note that $T_c=0$ at
half-filling \cite{Scalettar89}). Whereas the
kinetic energy is independent of the sign of $U$, the transverse current
response in Fig.~\ref{fig:resistivity-hubbard}(d) reveals a nonzero
superfluid density for $\oH^{c}_-$ that contributes to the sum
rule~(\ref{eq:fsumrule}). Because this component is not captured by
$\sigma'(\om)$,  the corresponding results for
$\rho$ based on estimator (3) are excluded from Fig.~\ref{fig:resistivity-hubbard}(f).

The resistivity, shown in Figs.~\ref{fig:resistivity-hubbard}(e) and (f), 
is thermally activated for attractive and repulsive
interactions at $T\gtrsim 1$. At lower temperatures, the attractive
(repulsive) model shows metallic (insulating) behavior. Comparison of
Fig.~\ref{fig:resistivity-hubbard}(f) with Fig.~\ref{fig:resistivity-fkm}(c)
reveals that $\rho(T)$ of the attractive Hubbard is qualitatively similar to 
the FKM results. Finally, the optical conductivity in
Figs.~\ref{fig:resistivity-hubbard}(g) and (h) has a Hubbard peak for both
signs of the interaction. Whereas the optical conductivity looks markedly
different for $U>0$ and $U<0$ at low temperatures, the filling-in
(suppression) at low frequencies in the repulsive (attractive) case
lead to almost identical results at high temperatures.

\section{Discussion}\label{sec:discussion}

In this section, we first summarize the mean-field theory that underlies the
identification of the strong-coupling metallic regime as an OM, before
discussing the similarities and differences in the present case. 
We then address the role of the gauge symmetry and quantum fluctuations,
and connections to other problems.

\subsection{Mean-field theory and orthogonal metals}\label{sec:model:mean-field}

Our discussion follows Ref.~\cite{PhysRevB.86.045128} but we use the notation
of Sec.~\ref{sec:slavespin}. More details and applications can be found in
Refs.~\cite{PhysRevB.81.155118,PhysRevB.86.045128,PhysRevLett.108.046401,PRYCHYNENKO201653}.
A product ansatz for the ground state of Hamiltonian~(\ref{eq:Hfs}) of the form
\begin{equation}\label{eq:meanfield-ansatz}
  \ket{\Phi}_\text{MF} = \ket{\phi}_f\otimes\ket{\phi}_s,\
\end{equation}
decouples the problem into a free-fermion part
\begin{equation}\label{eq:meanfield-fermions}
  \hat{H}_\text{MF}^{f} 
  = 
  -{t} \sum_{\las ij\ras,\sigma} g_{ij}
  ( \of^\dag_{i\sigma} \of^\nag_{j\sigma}  + \mathrm{H.c.})
%    \of^\dag_{j\sigma} \of^\nag_{i\sigma})
\end{equation}
and a transverse-field Ising model
\begin{equation}\label{eq:meanfield-ising}
  \hat{H}_\text{MF}^{s} 
  = 
  - t \sum_{\las ij\ras} J_{ij} 
  \os^z_i \os^z_j     
  -\frac{U}{4}\sum_i \os^x_i\,
\end{equation}
with the self-consistency conditions
\begin{equation}\label{eq:meanfieldparas}
  g_{ij} = \las \os^z_i \os^z_j\ras_s\,,\quad
  J_{ij} = \sum_\sigma \las ( \of^\dag_{i\sigma} \of^\nag_{j\sigma}  + \mathrm{H.c.}) \ras_f\,.
\end{equation}
In the simplest case, these expectation values are the
same for all nearest-neighbor bonds $\las ij\ras$. A more elaborate
mean-field approach
with a larger unit cell can be found, \eg, in Ref.~\cite{PhysRevLett.108.046401}.
Because the notation for the expectation values in
Eq.~(\ref{eq:meanfieldparas}) is specific to an ansatz such as
Eq.~(\ref{eq:meanfield-ansatz}), we drop the indices $s$ and $f$ from here on.

In the context of slave-spin theories, Eqs.~(\ref{eq:meanfield-fermions})--(\ref{eq:meanfield-ising})
have to be supplemented with local constraints
\cite{PhysRevB.81.155118,PhysRevB.86.045128,PhysRevB.86.045128}.
This is not the case if we consider Eq.~(\ref{eq:Hfs}) as the dual
(unconstrained slave-spin) representation of the FKM~(\ref{eq:HcQ}).

The quadratic fermionic part can be solved exactly for given $g_{ij}$ and
describes noninteracting fermions. The slave-spin part may be solved by
single-site or cluster mean-field methods
\cite{PhysRevB.81.155118}. Alternatively, more advanced schemes based on
saddle-point expansions or even numerical simulations are possible
\cite{PhysRevB.78.073108}. At any rate, the physics of the 2D quantum Ising
model~(\ref{eq:meanfield-ising}) is well understood.
Hamiltonian~(\ref{eq:meanfield-ising}) has a 
ferromagnetic (since $-t J_{ij}>0$) ground state with $\las \os^z_i\ras\neq
0$ for $U<U_c$ and a paramagnetic ground state with $\las \os^z_i\ras=0$ for
$U>U_c$, separated by a phase transition.

The effect of this phase transition on the original $c$ electrons becomes
apparent from their spectral function. The latter is a convolution of the
$f$-electron spectrum with that of the slave spins \cite{PhysRevB.86.045128},
\begin{equation}
  A(\bk,\omega) = \int\rmd\bq \int\rmd\Omega\,A^f(\bq,\Omega) A^s(\bk-\bq,\om-\Omega)\,.
\end{equation}
In the ordered phase, it takes the form \cite{PhysRevB.86.045128}
\begin{equation}\label{eq:akw-mean-field}
  A(\bk,\omega) = \las \os^z_i \ras^2 \delta(\omega-E_{\bk})\,.
\end{equation}
Here, the $\delta$ peak signals the absence of interactions in the
$f$-fermion sector and $\las \os^z_i \ras^2$ can be identified with the
$c$-fermion quasiparticle residue. The latter vanishes on approaching the
magnetic transition at $U_c$ and hence takes the role of a fermionic order parameter.
In the paramagnetic phase of the slave spins, $A^s$ and therefore also $A$ have a gap \cite{PhysRevB.86.045128}. 

As pointed out in Ref.~\cite{PhysRevB.86.045128}, earlier interpretations of
the slave-spin transition as a Mott transition of the $c$ fermions are in
general not correct. Unless vertex corrections can be neglected (as is the case
in infinite dimensions), a gap in $A(\bk,\omega)$ does not imply a
vanishing conductivity. Within the above mean-field theory, transport is fully
determined by the  noninteracting Hamiltonian~(\ref{eq:meanfield-fermions}).
The Drude weight ${D}\sim g_{ij} = \las \os^z_i \os^z_j\ras$
\cite{PhysRevB.86.045128} is nonzero  as long as the slave spins have
nonlocal correlations. It is only within local mean-field theories,
where $\las \os^z_i \os^z_j\ras=\las \os^z_i\ras^2$, that a vanishing
quasiparticle weight coincides with $D=0$ and hence insulating behavior.
Therefore, beyond local approximations, the gapped paramagnetic phase is
not a Mott insulator but a metallic state with a single-particle gap, namely
the OM~\cite{PhysRevB.86.045128}. Transport is fully carried by the
gapless $f$ electrons that may be regarded as being orthogonal to the gapped
$c$ fermions. This emerges naturally from the fact that the $f$ fermions
inherit both the physical spin [SU(2) symmetry] and the physical charge [U(1)
symmetry] of the original fermions \cite{PhysRevB.86.045128}. 

\subsection{Classification as an orthogonal metal}

The classification of the gapped regime at $T>T_Q$ and $U\gtrsim 6$
as an OM can be motivated from the duality between the FKM~(\ref{eq:HcQ}) and
the slave-spin representation~(\ref{eq:Hfs}) in conjunction with the
evolution of the slave-spin Green function in Fig.~\ref{fig:crossover-long}(d).
Whereas this connection requires some explanation with respect to the
absence of symmetry breaking beyond mean-field theory (see below), we also observe
the defining OM properties directly in the FKM representation: an absence
of quasiparticles [Figs.~\ref{fig:akw-z2}(d)--(f) and \ref{fig:nk-FSS}(b)]
together with a nonzero conductivity [Fig.~\ref{fig:crossover_dc}(b)], spin
and charge susceptibility [Fig.~\ref{fig:crossover-long}(d)]. They key role
of the single-particle spectral function in distinguishing FL and OM regimes
\cite{PhysRevB.86.045128} is fully borne out by our numerical results.

For weak $U$, the existence of well-defined quasiparticles
[Figs.~\ref{fig:akw-z2}(a)--(b) and \ref{fig:nk-FSS}(a)] is consistent with an FL,
for which the density of states $N(\EF)$ determines, \eg, the spin susceptibility
\cite{schofield1999non}. In contrast, for large $U$, the Drude peak in Fig.~\ref{fig:optcond_Z2_U}(b)
and the associated nonzero conductivity [Fig.~\ref{fig:crossover_dc}(b)]
cannot be reconciled with the absence of quasiparticles within FL theory.
Whereas the $f$ fermions remain
noninteracting in the models of Refs.~\cite{PhysRevB.86.045128,zhong2012correlated,chen2019metals},
the existence of orthogonal phases with residual interactions can readily be conceived.

In Sec.~\ref{sec:delineate-hubbard}, we showed that some properties of
the OM (single-particle gap, nonzero conductivity) are also realized in the
attractive Hubbard model at $T>T_c$ ($T_c=0$ for the case of half filling considered). The key distinction was the
existence of a gap for spin excitations in the latter, corresponding to a
paired FL or a pseudo-gapped metal. In contrast, the OM in the FKM has gapless
spin excitations, demonstrated in terms of the spin susceptibility
[Fig.~\ref{fig:crossover-long}(c)] and the dynamic spin structure factor
\RefI.  A spin gap gives rise to an entirely non-FL spin susceptibility
that decreases exponentially at low temperatures. Therefore, and given the
absence of $f$-fermion interactions that could produce such a gap in their models,
it is surprising that systems with uncondensed pairs of fermions are suggested
as candidates for OMs in Ref.~\cite{PhysRevB.86.045128}. In contrast,
our work appears to provide a faithful realization of the fermionic OM
deduced from mean-field theory.

Despite the close similarities with OMs in terms of gauge-invariant
properties, distinctions remain at the level of quantum numbers in the
unconstrained slave-spin formulation~(\ref{eq:Hfs}) of the FKM~(\ref{eq:HcQ}).
In the latter, the \ZII charge is conserved in space but not in time.
This should be compared with the exactly solvable fermion-spin models of
Refs.~\cite{PhysRevB.86.045128,zhong2012correlated} that have no local
symmetry and hence no associated \ZII charge. It is also different from 
$T=0$ numerical realizations of orthogonal (semi-)metals with spontaneously
generated constraints \cite{PhysRevX.6.041049,gazit2017emergent,arXiv:1804.01095,chen2019metals}.

To elucidate this point, we first note that according to
Eq.~(\ref{eq:hqcommute}), Hamiltonian~(\ref{eq:Hfs}) conserves the $\oQ_i$
and can therefore be represented in terms of blocks (\ZII charge or superselection sectors)
corresponding to fixed configurations of the $\oQ_i$
\cite{PhysRevB.96.205104}. Whereas the physical sector (\eg, $\oQ_i=1$ for
all $i$) is selected in constrained theories, all possible sectors contribute
to the physics observed at $T>T_Q$. 

\begin{figure}[t]
  \includegraphics[width=0.375\textwidth]{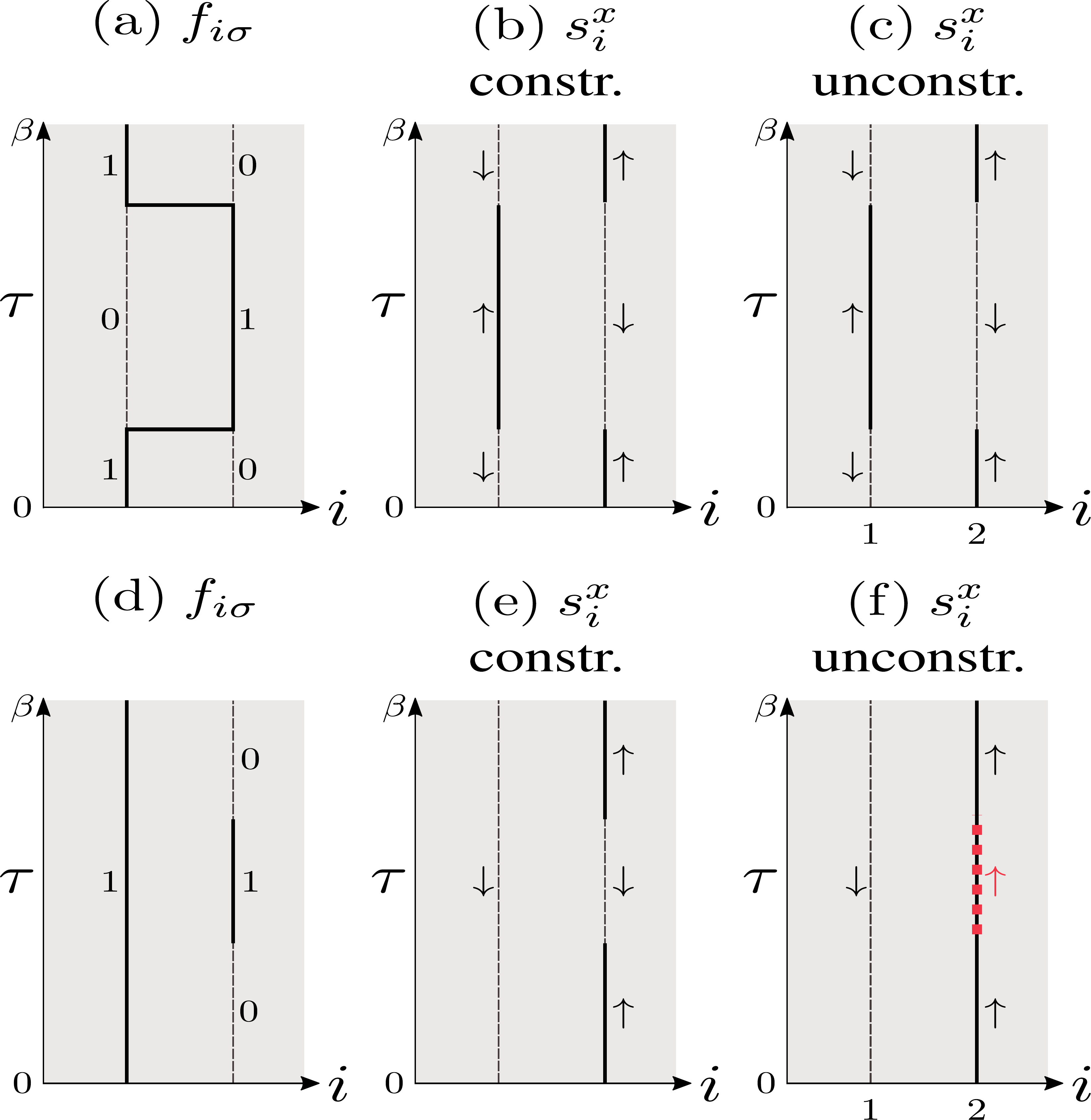} 
  \caption{\label{fig:worldline}
    World-line configurations of $f$ fermions and slave spins that illustrate
    the conservation or non-conservation of the \ZII charge under (a)--(c) hopping
    and (d)--(f) imaginary-time particle number fluctuations. (a) and (d)
    include occupation numbers without reference to the fermion spin $\sigma$.
  }
\end{figure}

If we only consider hopping processes, the \ZII charge remains conserved
because the hopping term in Eq.~(\ref{eq:Hfs}) changes both the fermion
parity and the slave-spin orientation at each of the sites involved.  This is
illustrated in terms of world lines of the $f$ fermions and the slave spins
in Figs.~\ref{fig:worldline}(a)--(c). Even though $\os^x_i (-1)^{\on_i}$ and
$\oQ_i$ are not linked at $T>T_Q$, they both retain their value along the
imaginary time axis. If constraints hold, all sites will have
the same value of $\oQ_i$ corresponding to the physical subspace chosen for
the slave-spin representation. Hence, with respect to hopping, constrained
and unconstrained theories are distinguished only by the values of the $\oQ_i$.
The latter have trivial flat world lines as a result of Eq.~(\ref{eq:hqcommute})
and are therefore not shown in Fig.~\ref{fig:worldline}. In contrast, the Ising degrees of freedom in
Eqs.~(\ref{eq:Hfs}) and~(\ref{eq:Hftau}) have a nontrivial
dynamics determined by the transverse-field term $\sim U$.

Constrained and unconstrained theories clearly differ with respect 
to local fluctuations of the occupation of the \ZII charged
$f$ fermions, as illustrated in Figs.~\ref{fig:worldline}(d)--(f).
In constrained theories, Gauss's law ensures that changes of the
$f$-fermion number entail corresponding changes of the slave spins to
preserve the eigenvalues of $\oQ_i$ [Fig.~\ref{fig:worldline}(e)]. In
contrast, without constraints, the slave-spin configuration remains
unchanged in Fig.~\ref{fig:worldline}(f), mimicking a transition between
different charge sectors as a function of imaginary time and corresponding to
a violation of \ZII charge conservation.

\subsection{Gauge-invariant picture}\label{sec:gauge-symmetry}

The above mean-field theory of the FL-OM transition provides a
conceptual framework to understand our numerical findings. In particular, 
the significantly different behavior of the slave spins at weak and strong
coupling justifies the fractionalization ansatz~(\ref{eq:slave-spin1}). At the same time, the mean-field
results require some clarification in the context of gauge
invariance and our numerical simulations. Although the connections between
the FKM~(\ref{eq:HcQ}) and the Hubbard model are of significant interest, we
point out that a constrained slave-spin representation of the former is not
desirable because it would produce known Hubbard results at any temperature,
which are more easily accessible in the representation~(\ref{eq:hubbard}).

To be concrete, we focus on the repulsive Hubbard model, even though our
considerations apply more generally. The theory of Sec.~\ref{sec:model:mean-field} suggests a $T=0$ FL-OM transition associated
with a phase transition of the slave Ising spins driven by the transverse
field $h=U/4$. The ferromagnetic phase corresponds to the FL, the
paramagnetic phase to the OM. In contrast, the true ground state of the 2D
Hubbard model and the FKM~(\ref{eq:HcQ}) (equivalent to the Hubbard model at
$T=0$) is known to be an antiferromagnetic Mott insulator for any $U>0$. 
The paramagnetic phase of the mean-field description reduces to a Mott
insulator in the local approximation, but the theory does not capture the magnetic order
or the weak-coupling instability due to perfect nesting. Clearly, mean-field slave-spin
approaches have rather limited predictive power for specific models. 

As discussed in Sec.~\ref{sec:model:symmetries}, a gauge redundancy
(constrained theories) or local symmetry (unconstrained theories), inherent
to the slave-spin representation of Eq.~(\ref{eq:Hfs}), implies 
that expectation values of gauge-dependent operators vanish.  Elitzur's
theorem \cite{PhysRevD.12.3978} generally rules out a spontaneous breaking of 
gauge invariance or local symmetries in finite dimensions.

The local \ZII transformations are generated by
the $\oQ_i$. In an exact slave-spin treatment (constrained gauge theory, or unconstrained theory
below $T_Q$), they can be applied at any point in spacetime
\cite{RevModPhys.51.659}. This implies
\begin{itemize}
\item[(i)] the absence of a phase transition of the
slave spins characterized by a local order parameter,
\item[(ii)] the absence of any nonlocal spatial or temporal correlations
of $f$ fermions or slave spins.
\end{itemize}

For constrained and unconstrained theories, the fact that $\os^z_i$ and
$\of^{(\dag)}_{i\sigma}$ carry a gauge charge requires that the magnetization
$m=\las \os^z_i\ras$, the mean-field Drude weight $D\sim \las \os^z_i
\os^z_j\ras$, and $\las \of^\dag_{i\sigma} \of^\nag_{j\sigma} \ras$ 
vanish. For example, using Eq.~(\ref{eq:fandtaulocal}) directly yields 
\begin{equation}
  m = \las s^z_i \ras = \las \oQ_i s^z_i \oQ_i \ras = - \las s^z_i \ras = 0\,.
\end{equation}
For the Hubbard model, imposing these restrictions on
the slave-spin mean-field theory would yield a Mott insulator for any $U>0$.

A more appropriate interpretation of the mean-field results comes from lattice gauge theories, 
where predictions based on a forbidden local order parameter
are often in good agreement with more advanced gauge-invariant approaches
\cite{frohlich1981higgs,drouffe1983strong}. 
For example, the ferromagnetic order of the slave spins in the FL regime  may be regarded
as obtained for a fixed gauge [the product ansatz~(\ref{eq:meanfield-ansatz})
reduces the local gauge symmetry to a global \ZII symmetry]. If the different
saddle points related by gauge transformations make similar contributions to
gauge-invariant quantities, averaging will yield nontrivial results while
restoring gauge invariance. The symmetry-forbidden expectation values are
hence merely the mean-field encoding of gauge-invariant physics.
More generally, the physics extracted by
gauge-fixing can also be observed in a gauge-invariant formulation \cite{frohlich1981higgs,drouffe1983strong}. Here, 
 the unconstrained and hence not fully
gauge-invariant representation~(\ref{eq:Hfs}) has a dual, manifestly
gauge-invariant representation in terms of the FKM~(\ref{eq:HcQ}).
Similarly, OMs can be observed in exactly solvable models that do not have 
a local \ZII symmetry \cite{PhysRevB.86.045128,zhong2012correlated} as well
as in numerical simulations of gauge theories coupled to matter with
spontaneously generated constraints \cite{arXiv:1804.01095,chen2019metals}.

For the FKM~(\ref{eq:HcQ}), as well as other unconstrained gauge theories,
nontrivial imaginary-time
correlations of gauge-dependent operators are explicitly allowed and observed
because point (ii) above only applies to spatial correlations. For example,
the temporal freezing of the slave spins at $U=0$ can mimic long-range
order on the relevant time scales while preserving a vanishing (spatial)
magnetization. An adiabatic connection between $h=0$ and
$h>0$ has been established for the Higgs phase of Ising lattice gauge
theories \cite{PhysRevD.19.3682}. 

The crossover from frozen
to disordered slave spins tracks the opening of the single-particle gap and
the associated breakdown of a quasiparticle description. (This is opposite to
non-FL physics arising {\em from} spin-freezing \cite{PhysRevLett.101.166405}.)
However, it does not account for the pronounced differences between
paramagnetic Mott and OM phases. Whereas the former is a fully
gauge-invariant state determined by local physics, the latter crucially
relies on vertex corrections, encoded at the mean-field level in the
nonvanishing spatial correlators $\las \os^z_i \os^z_j\ras$ and $\las
\of^\dag_{i\sigma} \of^\nag_{j\sigma} \ras$. Such physics is fully captured
by our numerical simulations, but absent in the DMFT limit
$\text{D}=\infty$ where the OM is replaced by a Mott insulator.
For $\text{D}=\infty$, both spatial and temporal correlations are
allowed by Elitzur's theorem \cite{maslanka1988symmetry}. However, spatial
correlations and vertex corrections are
generically absent, resulting in a strong
time-space asymmetry of gauge-dependent correlators similar to the 2D FKM~(\ref{eq:HcQ}).
Although vertex corrections may be
expected to be less important at high temperatures, DMFT does not become
exact for $T\to \infty$ \cite{PhysRevB.94.235115}. This is consistent with our
observation of an OM at high temperatures but its absence in DMFT.

An interesting question is whether the FL and OM regimes are connected by a
phase transition or a crossover. Whereas the mean-field scenario with the local order
parameter $\las \os^z_i\ras$ violates gauge invariance, it is known 
that in $2+1$ dimensions the Ising lattice gauge theory is dual to the
standard Ising model with a global \ZII symmetry \cite{wegner1971duality}. 
The phase transition of the latter is encoded in the former in terms of the
different behavior (area vs. perimeter law) of the
Wegner-Wilson loop correlator
\cite{wegner1971duality,RevModPhys.51.659,FradkinBook}, a nonlocal
gauge-invariant order parameter. Whether such a transition survives for
nonzero matter coupling and at finite temperature cannot be answered in
general. For example, a 3+1D Ising theory coupled to a Higgs field exhibits a
first-order transition (crossover) below (above) a critical Higgs mass
\cite{PhysRevLett.77.2887}. Our numerical data appear consistent with an
FL-OM crossover driven by the disordering of the slave-spins and with a
smooth evolution of gauge-invariant properties. 
A continuous
FL-OM quantum phase transition, characterized by a nonlocal order parameter,
 was numerically realized for a \ZII gauge theory in
 Ref.~\cite{chen2019metals}.

A phase transition of the slave spins with a local order parameter is 
observed at $T=0$ for $\text{D}=\infty$ \cite{PhysRevB.91.245130},  where it
is not forbidden by Elitzur's theorem \cite{maslanka1988symmetry}. Because the constraints are
irrelevant, the paramagnetic DMFT solution of the unconstrained slave-spin
theory gives exact results and $\las \os^z_i\ras$ serves as an order parameter
\cite{PhysRevB.91.245130}. However, in the absence
of vertex corrections, the OM is replaced by a Mott insulator, see
Fig.~\ref{fig:grandphasediagram}. Even for $\text{D}=\infty$, the mean-field
picture apparently breaks down at $T>0$, where DMFT suggests a vanishing slave-spin
magnetization for all $U$. As pointed out in Ref.~\cite{PhysRevB.91.245130},
this is consistent with an identification of the slave-spin magnetization
with the quasiparticle weight, the latter being defined only at $T=0$.
In DMFT, the first-order Mott transition in Fig.~\ref{fig:grandphasediagram} is
instead characterized by the opening of a gap in the slave-spin spectrum
\cite{PhysRevB.91.245130}, quite similar to the disordering crossover
observed here (frozen slave spins imply gapless excitations).

While the slave-spin magnetization is gauge-dependent and hence zero for
$\text{D}<\infty$, mean-field theory establishes a link to the
gauge-invariant quasiparticle residue. The latter is given by the jump of
$n(\bk)$ at $T=0$ but vanishes for $T>0$. A more general gauge-invariant
order parameter is $\las \os^x_i \ras$; $\os^x_i$ may be regarded as a
disorder operator \cite{PhysRevB.3.3918} whose condensation amounts to a
proliferation of domain walls in imaginary time and hence a disordering of
the slave spins. The data for the slave-spin Green function in
Fig.~\ref{fig:crossover-long}(d) can qualitatively be understood as a crossover
from $\las \os^x_i \ras=0$  at small $U$ to $\las \os^x_i \ras>0$ at large $U$.

For $\text{D}=\infty$, or in constrained slave-spin theories, $\las\os^x_i\ras$ is
directly related to the fermion double occupancy $\las \on_{i\UP}
\on_{i\DO}\ras$, see Eq.~(\ref{eq:docctotransversefield}). The latter is manifestly
gauge invariant and provides an experimentally
accessible \cite{jordens2008mott} Ising order parameter for the Mott
transition also at $T>0$ \cite{Brinkman70,PhysRev.137.A1726,PhysRevLett.83.3498,kim2014estimate}.
The O(4) symmetry of the FKM~(\ref{eq:HcQ}) implies that the double occupancy
is independent of $U$, corresponding to an average over the values for the
attractive and the repulsive Hubbard model. However, it is not clear if this
should be interpreted as evidence for a crossover rather than a phase
transition, as done in Ref.~\cite{tran2018fractionalization}. A crossover
scenario is favored by the smooth evolution and absence of noticeable size
effects in Fig.~\ref{fig:compress-vs-U}, as opposed to the clear signature of
the magnetic phase transition (FL to Hubbard regime) in
Fig.~\ref{fig:compress-fkm}(b). Although the OM is distinguished from
a Mott state by nonlocal effects, the generically weak
finite-size effects in Figs.~\ref{fig:nk-FSS}(b),
\ref{fig:crossover_dc}(c), \ref{fig:crossover_dc}(d),
and~\ref{fig:compress-vs-U} suggest predominantly short-range correlations.

\subsection{\ZII fractionalization and topological order}\label{sec:discussion:qshstar}

Slave-spin mean-field theories have also been applied to models of
interacting fermions with topologically nontrivial band structures.
Such approaches yield \ZII-fractionalized quantum spin Hall (QSH) and Chern
insulators \cite{PhysRevLett.108.046401,PRYCHYNENKO201653}. Here, we focus
on the so-called QSH* phase reported for a Hubbard model with
spin-orbit coupling~\cite{PhysRevLett.108.046401}.  It is separated from the
regular QSH phase by an Ising transition of the slave spins. In the
QSH* phase, where the slave spins are disordered, the characteristic helical
edge states \cite{KaMe05a} are gapped in terms of the $c$ fermions but
remain metallic with respect to the $f$ fermions. The prospect of
such a phase---absent in unbiased simulations of the Kane-Mele-Hubbard
model \cite{PhysRevB.90.085146}---partially motivated our work.

The classification of the QSH* phase as
topologically ordered \cite{PhysRevB.40.7387} (\ie, with a four-fold
degenerate groundstate on a torus and fractional excitations) is related to
the existence of $\pi$ fluxes in the mean-field solution after transforming
from site to bond Ising variables and allowing for fluctuations of the latter
\cite{PhysRevLett.108.046401}. Specifically, in the QSH* phase, products 
of the hopping integrals $g_{ij}\sim\las \os^z_i \os^z_j\ras$ over closed paths equal $-1$.

It is interesting to consider these findings in the context of our
simulations. For any fixed slave-spin configuration $\mathbf{s}$, closed
paths on a bipartite lattice always consist of an even number of bonds, see
Fig.~\ref{fig:isingspins}. Given the exact mapping between site and bond spins,
each slave spin $\os^z_i$ affects the sign of all bond spins $\hat{\xi}^z_{ij}$ associated with site
$i$. Flipping a slave spin that is part of a closed loop $\square$ changes the sign of
two bond spins and hence leaves the total sign of $\prod_\square g_{ij}$
unchanged; bond spins cannot be flipped independently. The absence of fluxes
for each configuration also implies their absence on average. None of the
saddle points that play a role in our simulations therefore have $\pi$ fluxes.
This reasoning applies to the square lattice studied here and to the  honeycomb lattice considered in
Ref.~\cite{PhysRevLett.108.046401}. Importantly, these conclusions are
completely independent of whether or not the slave-spin constraints hold.

In the slave-spin context, the possibility of $\pi$ fluxes and hence
topological order, as observed in mean-field theory \cite{PhysRevB.40.7387},
only arises if the $\hat{\xi}^z_{ij}$ are treated as independent degrees of
freedom. Thereby, the exact duality between site and bond variables is lost
and the relevance of the theory for its starting point---the microscopic
Hamiltonian---is no longer obvious. For example, a QSH* phase has not been
observed in QMC simulations of Kane-Mele-Hubbard and Kane-Mele-Coulomb models \cite{PhysRevB.90.085146}.
The interesting $\pi$-flux solutions may be interpreted as an effective mean-field description of
physics that may manifest itself in a different way in exact
simulations. Similar to our FKM, where we observe OM behavior even though the
relevant mean-field expectation values vanish as a result of gauge invariance,
a QSH* phase should have a four-fold degenerate ground state and a
characteristic response to externally imposed $\pi$ fluxes \cite{PhysRevLett.108.046401}.

\subsection{\ZII gauge theories and designer Hamiltonians}\label{sec:zii-gauge-theories}

Models of fermions coupled to Ising spins  have recently been studied in
Refs.~\cite{PhysRevX.6.041049,gazit2017emergent,arXiv:1804.01095,PhysRevLett.121.090402,gonzalez2019intertwined}.
For relations between FKMs and lattice-gauge theories, see Ref.~\cite{PhysRevB.96.205104}.
There has also been significant interest in experimental realizations of
lattice gauge theories using cold atoms in optical lattices
\cite{zohar2015quantum,smith2018dynamics,barbiero2018coupling}.

Hamiltonian~(\ref{eq:Hftau}) is formally identical to the $N=2$ model of
Refs.~\cite{PhysRevX.6.041049,gazit2017emergent,arXiv:1804.01095}. These
works also do not impose Gauss's law but focus on $T=0$ where the constraints
are spontaneously generated. A fundamental 
difference is that here the bond spins in Eq.~(\ref{eq:Hftau}) are defined as
products of slave spins.  Accordingly, the bond-spin
representation contains a star-operator term in Eq.~(\ref{eq:Hftau}) but 
a simpler transverse field term in Refs.~\cite{PhysRevX.6.041049,gazit2017emergent,arXiv:1804.01095}.
This rules out $\pi$-flux configurations that underlie
the Dirac-fermion phases of
Refs.~\cite{PhysRevX.6.041049,gazit2017emergent}. Nontrivial flux
configurations in unconstrained gauge theories are also discussed in
Ref.~\cite{PhysRevB.96.205104}, whereas they are naturally absent in
related work for 1D systems
\cite{PhysRevLett.118.266601,PhysRevLett.121.090402,gonzalez2019intertwined}.

A quantum phase transition between conventional and orthogonal FL phases was
recently studied numerically in Ref.~\cite{chen2019metals}. The corresponding
Hamiltonian describes $f$ fermions,  slave spins, and \ZII gauge
fields. However, in contrast to slave-spin theories such as Eq.~(\ref{eq:Hfs}).
there is no explicit hopping term for the gauge-invariant $c$ fermions. For
the parameters considered, $\pi$ fluxes are absent. The QMC results capture
the nonlocal order parameter that characterizes the FL-OM transition
in a fully gauge-invariant setting. 

\subsection{Localization, boson-controlled hopping} 

It is also interesting to compare our findings for the FKM~(\ref{eq:HcQ}), with its dual
representation~(\ref{eq:Hfs}), to models for
localization without disorder. The latter can be formulated either as
unconstrained \ZII lattice gauge theories of spinless fermions or as spinless
FKMs. As mentioned in Sec.~\ref{sec:struct-phase-diagr}, the half-filled
spinless FKM (same number of $c$ fermions and $l$ fermions, $L^2$ particles
in total) with interaction $U \sum_i \on_i \on_i^{l}\sim U \sum_i \on_i
\oQ_i$ exhibits CDW order for $T<T_c$ \cite{RevModPhys.75.1333}. Earlier work
\cite{PhysRevB.74.035109} suggested identical properties for any $U>0$ in the
disordered region at $T>T_c$, with an FL only at $U=0$ \cite{PhysRevB.46.1261}.
More recently, a crossover between an Anderson insulator at weak $U$ and
a Mott insulator at large $U$
\cite{PhysRevLett.117.146601,oliveira2018classical} was revealed for the
infinite system. A crossover between a bad metal and an Anderson
insulator takes place as a function of system size \cite{PhysRevLett.117.146601}.
The topology of the phase diagram is remarkably
similar to Fig.~\ref{fig:phasediagram}. We expect the apparently robust metallic
behavior observed here to be associated with the fundamentally
different nature of the coupling between itinerant and localized
fermions. For a fixed configuration of the $\oQ_i$, the spinless FKM
describes the impact of a random onsite potential,
whereas our spinful FKM describes a Hubbard interaction with a random sign.
The time-dependent dynamics of the 1D spinless FKM in the dual lattice-gauge
representation also reveals localization without explicit
disorder~\cite{PhysRevLett.118.266601}. Connections between fermions with
binary disorder (corresponding to configurations of the Ising constraints
$\oQ_i$) and unconstrained gauge theories are discussed in Ref.~\cite{smith2018dynamics}.

The disorder of the slave spins in our model determines the coherent
motion of single $c$ fermions. For $U=\infty$, the slave
spins form a perfect paramagnet, $\ket{\psi}=\prod_i
(\ket{\UP}^s_i+\ket{\DO}^s_i)/\sqrt{2}$, and $\bra{\psi} \hat{H}_0^{fs}
\ket{\psi}=0$. The strong reduction of the conductivity in Fig.~\ref{fig:crossover_dc}
may be attributed to enhanced scattering at large $U$ and suggests possible connections
to other models of fermions coupled to bosonic excitations of a background
medium \cite{ed06,alvermann:056602,sous2019fractons}. Non-quasiparticle
transport was studied in Ref.~\cite{werman2017non} using a model of SU($N$)
fermions coupled to classical bond phonons, whereas the mapping of Eq.~(\ref{eq:Hfs})
to a quantum Su-Schrieffer-Heeger model was discussed in Ref.~\cite{PhysRevX.6.041049}.

\section{Conclusions}\label{sec:conclusions}

We carried out a comprehensive study of a recently introduced
FKM with an OM regime at strong interactions and above a critical
temperature. It is related to the 2D Hubbard model in the
(unconstrained) slave-spin representation via an exact duality.
 We demonstrated the quantitative agreement with the Hubbard
model in several limits. The pronounced differences between weak and strong
coupling in the high-temperature phase were further established by
diagnostics such as the momentum distribution function, different estimators
of the conductivity, and the optical conductivity. These results underline
the existence of an OM regime characterized by the absence of
quasiparticles but robust metallic behavior. Superconductivity and pairing
were explicitly ruled out. We also addressed the temperature dependence in
the weak-- and strong-coupling regimes, observing eventual saturation of the resistivity, in
contrast to recent work on the doped Hubbard model \cite{huang2018strange}. An
explicit comparison to half-filled attractive and repulsive Hubbard models
was presented and revealed clear differences.

We also addressed difficulties in reconciling the mean-field
slave-spin results with gauge invariance. While the OM concept has been corroborated
in exactly solvable models \cite{PhysRevB.86.045128} and numerical simulations of a designer
lattice-gauge theory, our model~(\ref{eq:HcQ}) provides a realization (albeit
only at $T>T_Q$) in a rather simple microscopic model defined purely in terms
of gauge-invariant objects and closely related to the Hubbard model. Whereas
we focused on the slave-spin picture, it appears worthwhile to understand the
OM physics directly in terms of the FKM. This includes the role of (domain
walls between) extended attractive and repulsive Hubbard domains formed by
the $\oQ_i$ at $T\gtrsim T_Q$ for the Drude type contribution in
Fig.~\ref{fig:optcond_Z2_U}(b) and the spin-wave signatures in
Fig.~\ref{fig:akw-z2}(d) [reminiscent of those in Fig.~\ref{fig:resistivity-hubbard}(a)].

An interesting extension of our work are Dirac
fermions on the honeycomb or $\pi$-flux lattice. The
corresponding Hubbard model has an intriguing fermionic quantum critical
point \cite{Herbut09a,Assaad13,Toldin14,Otsuka16}.
Doping should act like a magnetic field for the constraints $\oQ_i$
and we expect the physics of the attractive Hubbard model to emerge.
Another fruitful direction are SU($N$) fermions.  Finally, returning to the
schematic, extended phase diagram of Fig.~\ref{fig:grandphasediagram}, it
would be fascinating to understand how the 2D Hubbard model evolves
as a function of $\text{D}$. The FL-OM crossover in our 2D simulations may be
interpreted as a finite-dimensional analog of the
$\text{D}=\infty$ crossover from FL to Mott insulator in the paramagnetic
DMFT solution. This raises the question
if the critical end point of the latter has a counterpart in a finite
dimension $\text{D}>2$, leading to the FL-OM phase transition anticipated
in Ref.~\cite{PhysRevB.86.045128} and observed in Ref.~\cite{chen2019metals}.

Finally, experimental advances with cold atoms in optical lattices may allow
for a realization of the FKM
\cite{lewenstein2007ultracold,esslinger2010fermi,mazurenko2016experimental,gall2019simulating}. In
particular, the high-temperature OM appears less challenging to detect than, for
example, antiferromagnetic ground states.

\vspace*{-1em}
\begin{acknowledgments}
\vspace*{-1em}
We are grateful to G. Czycholl, M. Fabrizio, J. Freericks,
S. Kirchner, J. Knolle, and A. R\"uegg for helpful discussions.
MH was supported by the DFG through SFB 1170 ToCoTronics. FFA was supported
by the DFG through the
W\"urzburg-Dresden Cluster of Excellence on Complexity and Topology in
Quantum Matter ct.qmat (EXC 2147, project-id 39085490).
We further thank the John von Neumann Institute for Computing (NIC) for computer resources on the
JURECA~\cite{Juelich} machine at the J\"ulich Supercomputing Centre (JSC).
\end{acknowledgments}

%\bibliographystyle{apsrev4-1}
%\bibliography{../refs}


\end{document}